\documentclass[jkps,showpacs,showkeys,twocolumn]{revtex4-1}

\usepackage{graphicx}
\usepackage{amssymb}
\usepackage{amsmath}
\usepackage{bm}

\usepackage{subcaption}
\usepackage[font=small]{caption}
\usepackage{multirow}
\usepackage{dcolumn}

\begin{document}

\setcounter{page}{0}
\title{Quadratic Density Response Function of a Two Dimensional Electron Gas}

\author{Chang-Jin Lee}
\email{changjin21c@daum.net}
\affiliation{Donostia International Physics Center (DIPC), 20018 San Sebastian, Spain}

\date{\today}

\begin{abstract}
A closed-form expression of the retarded quadratic density response function of a two dimensional non-relativistic electron gas at zero temperature is written in terms of a non-analytic complex function. A careful analysis is made of the mathematical mechanism of exact static features such as vanishing, discontinuity and peaks. Special attention is paid to an ambiguity in treating the phase of the non-analytic complex function.
\end{abstract}



\maketitle

\section{Introduction}

In the response theory based on an electron gas model, an essential role has been played by the retarded $\ell$inear Density Response Function ($\ell$DRF) ever since its specific form of the 3-Dimensional Electron Gas (3DEG) was known by Lindhard while the 2D $\ell$DRF by Stern\cite{Lin54,Stern67}. 
However, the higher order DRFs become important when a nonlinear effect comes into a play. A prominent example is the so-called Barkas effect in the electronic energy loss problem\cite{BBS56,FT47}, which could not be explained by the linear density response theory\cite{LW64}. 

On the way to tackle the Barkas effect, physicists tried to obtain the analytic form of the quadratic order DRF. The low-frequency limit of the real part of 3D qDRF was obtained by Lloyd and Sholl \cite{LS68} whereas the same limit of the imaginary part was done by Hu and Zaremba\cite{HZ88}. However, the complete form of the wavevector- and frequency-dependent 3D qDRF did not appear until Cenni and Sarraco(CS) reported its real part\cite{CS88,CCCS92} while Pitarke, Ritchie, Echenique and Zaremba reported its imaginary part\cite{PREZ93}. Soon after the work of Pitarke \textit{et al.}\cite{PREZ93}, Richardson and Ashcroft(RA) reported another form of the imaginary part obtained by a technique based on the Feynman trick in the imaginary frequency space\cite{RA94}.  

A specific form of both the time-ordered and the retarded qDRF in 3D was reported by Pitarke, Ritchie and Echenique(PRE)\cite{PRE95}, and later by Del Rio and Pitarke\cite{GP01}. They followed the field-theoretic approach, and adopted the real part of 3D qDRF obtained by CS\cite{CCCS92}. They also found an interesting collinear-limit expression that looks like the \textit{decomposition theorem} claimed by CS. As for the imaginary part, they took a different approach from RA. They separated the contribution of the infinitesimal imaginary frequencies from the principal part. The same method was applied to the 2D qDRF by Bergara, Pitarke and Echenique(BPE)\cite{BPE99}. Before BPE came out, Zhang also reported a closed-form expression of the static 2D qDRF and even the expression of the static 2D DRF generalized up to the infinite order\cite{Z91}.

On the other hand, Rommel and Kalman obtained other closed-form expressions of the wavevector- and frequency-dependent retarded qDRFs in various dimensions without relying on the so-called Feynman trick. They calculated the real and imaginary parts all at once not separately. They could write the 2D and 3D qDRFs in terms of a complex function that looks much simpler than any other reported expression. Especially in a 2DEG they noticed a number of conspicuous remarkable static features that will be referred to as \textit{vanishing, discontinuity} and \textit{peaks}\cite{RKG98,R99,KR00a, KR00b}.  However, there had been some mistakes in their calculations that became clarified and corrected later\cite{Lee08a}.

The main difficulties in obtaining the closed-form of the qDRF are two fold. One comes from the intricacy involved in the derivation of the qDRF in relation to the three-point density correlation functions\cite{HZ88,BCPE97,GP01,NN02}, or more generally the nonlinear fluctuation-dissipation theorem\cite{WH02}. The other comes from the different mathematical techniques involved in the integration of the one-loop triangle diagrams of the qDRF\cite{Z91,RA94,PRE95,RKG98,R99,BPE99}. The closed-form expression of the qDRF varies depending on the integration technique that has been adopted, making the comparison hard to be achieved.

However, there are a number of things that can be said about the complex closed form(Eq.(\ref{chi in mu}) and Eq.(\ref{F function single})) of the 2D qDRF presented in this paper. 
Firstly, the retarded expression(Eq.(\ref{F}) and Table \ref{Ff}) in this paper is valid for a 2DEG only at zero temperature.
Secondly, the retarded expression(Eq.(\ref{F}) and Table \ref{Ff}) in this paper is identical to the Eq.(2.16) in BPE with no difference except the overall sign. In fact, BPE mentioned that Eq.(2.16) is retarded\cite{FN22}. 
Thirdly, the complex closed form of the 2D qDRF is different from the 2D qDRF in BPE even if they are composed of the similar expressions in part. The complex closed form in this paper is written in terms of a non-analytic complex function, and its real and imaginary parts are not separated but given in terms of a single complex function. In fact, there is a difficulty in extracting the real part from the complex closed form due to the \textit{phase ambiguity} that will be clearer in this paper.  On the other hand, the specific forms of the real and imaginary parts in BPE are written in the separate forms. 
Fourthly, the complex closed form satisfies the well-known symmetry of the retarded qDRF such as the wavevector inversion symmetry and the reality condition in a real space\cite{FN23}. In addition, it behaves in a way as one expect in the static long-wavelength limit as well as in the collinear limit. 
Finally, one may find that the complex closed-form expression suffers from the ambiguity(or defect) in treating the phase of a complex function\cite{FN24}. However, the ambiguity is not due to the weakness of the complex closed form itself but it arises from the ignorance of the weakness of the current computational algorithm: especially in treating the phase of the non-analytic complex function.

The plan of this paper is as follows. In Sec.\ref{2DqDRF}, a closed-form expression of the wavevector- and frequency-dependent retarded qDRF of a 2DEG will be written in terms of a complex function that will be analytically calculated with a mathematical rigor. In Sec.\ref{ch static}, a careful analysis will be made of the mathematical mechanism of vanishing, discontinuity and peaks with emphasis on the exact static 2D qDRF. In Subsec.\ref{ambiguity}, special attention will be paid to an ambiguity in treating the phase of the non-analytic complex function. The conclusion will be given in Sec.\ref{conc}.

\section{\label{2DqDRF}2D quadratic density response function}

Let us focus on a homogeneous non-interacting electron gas at zero temperature driven by a classically behaving external potential $\Phi(\mathbf{r},t)$. The Hamiltonian $H(t)$ consists of the kinetic part $H_0$ and the interaction part $H_e(t)$: $H(t) = H_0 + H_e(t)$. With the form of $ \hat{H}_e (t) = \int \mathrm{d} \mathbf{r} \, \rho (\mathbf{r},t) \Phi(\mathbf{r},t)$ as the interaction part, we assume that the quadratic order induced electron density $\langle \rho(\mathbf{r},t) \rangle_0^\mathrm{2}$ may be written in the symmetric form of\cite{FN25}
\begin{eqnarray}
&&   \big{\langle} \hat{\rho}(\mathbf{r},t) \big{\rangle}^{(2)}_0 \! \nonumber \\
  && = \!
  \frac{1}{2!} \left( \frac{-i}{\hbar} \right)^2    \!\! 
      \int_{-\infty}^\infty \!\! \!\! \!\! \mathrm{d} t_1 \;  \theta(t - t_1) 
      \int_{-\infty}^\infty \!\! \!\! \!\! \mathrm{d} t_2 \;  \theta(t - t_2)  \nonumber \\
&& \hspace{0.2cm} \times
      \Big\{ 
	\theta(t_1 - t_2)  \big{\langle}  \Psi_0  \big{|} \big[ [ \hat{\rho}(\mathbf{r},t),  \hat{H}_e(t_1)], \;   
          \hat{H}_e(t_2)  \big] \big{|} \Psi_0   \big{\rangle}  \nonumber \\
&& \hspace{0.5cm}  
          + \theta(t_2 - t_1)   \big{\langle}   \Psi_0 \big{|} \big[ [ \hat{\rho}(\mathbf{r},t),  \hat{H}_e(t_2)], \;   
          \hat{H}_e(t_1)  \big] \big{|} \Psi_0   \big{\rangle} 
	\Big\}  , \label{rho2}
\end{eqnarray}
where the bracket $[\hat{\rho}, \hat{H}_e]$ stands for the commutator and $|\Psi_0 \rangle$ is the unperturbed ground state of an electron gas: simply a Slater determinant.
It is noteworthy that the quadratic order density response is retarded due to the condition of the Heaviside step function $\theta(t)$ that is defined as $\theta(t) = 1$ for $t \ge 0$ and zero for $t<0$. 

By performing the Fourier transform of Eq.(\ref{rho2}), one will end up with
\begin{eqnarray}
  \! \! \langle \hat{\rho}(\mathbf{k}, \omega) \rangle^{(2)}_0
   \!\! & = & \!\!
  \frac{1}{\mathcal{V}} \sum_{\mathbf{k}_1} \int \frac{\mathrm{d} \omega_1}{2\pi}
  \frac{1}{\mathcal{V}} \sum_{\mathbf{k}_2} \int \frac{\mathrm{d} \omega_2}{2\pi} \nonumber \\
&& \times \;
  \mathcal{V} \delta_{\mathbf{k}_1 + \mathbf{k}_2 - \mathbf{k}}  (2\pi) \delta(\omega_1
  + \omega_2 - \omega) \nonumber \\
&& \times \;
  \chi(\mathbf{k}_1, \omega_1 ; \mathbf{k}_2,\omega_2)
  \Phi(\mathbf{k}_1, \omega_1) \Phi(\mathbf{k}_2, \omega_2), \label{rho k space}
\end{eqnarray}
where $ \chi(\mathbf{k}_1, \omega_1 ; \mathbf{k}_2,\omega_2)$ stands for the retarded qDRF of a non-interacting electron gas. 
Even if $\chi^0$ is the usual notation for the DRF of a non-interacting electron gas, $\chi$ will be used for it throughout this paper: refer to the appendix \ref{app qDRF} for more details about $\chi$.

The retarded qDRF of $ \chi(\mathbf{k}_1, \omega_1 ; \mathbf{k}_2,\omega_2)$ can be
given as a summation of six $F(asb) \equiv F(\mathbf{k}_a, \omega_a ; \mathbf{k}_b, \omega_b)$ functions, which are defined as
\begin{equation}
F(asb)  
=
\frac{1}{\mathcal{V}} \sum_{\mathbf{q},\sigma} n_{\mathbf{q}} \frac{1}{\omega_a   - \bar{\omega}_\mathbf{q}(\mathbf{k}_a : 0)} \cdot \frac{1}{\omega_b   - \bar{\omega}_\mathbf{q}(\mathbf{k}_b : 0)},   \label{F}
\end{equation}
where $\sigma$ is the spin index and $\mathcal{V}$ is the volume factor. $\omega_a \equiv \pm ( \omega_a' + i \delta )$ is a complex value with the real $\omega_a'$$(a = 0,1,2)$ and the positive infinitesimal $\delta$.  $\bar{\omega}_\mathbf{q} (\mathbf{k}_a : \mathbf{k}_b)$ is defined as $\hbar \bar{\omega}_\mathbf{q} (\mathbf{k}_a : \mathbf{k}_b) \equiv \hbar^2 (\mathbf{q} + \mathbf{k}_a)^2/2m_e - \hbar^2 (\mathbf{q} + \mathbf{k}_b)^2/2m_e$.  $(asb)$ is a symbol denoting one of the permutations of $(012) \equiv (\mathbf{k}_0, \omega_0; \mathbf{k}_1, \omega_1; \mathbf{k}_2, \omega_2)$, which indicates that the qDRF is closely related to the permutation of variables in $F(asb)$'s.  The variables of the six $F(asb)$'s are specified in Table \ref{Ff}, where $\omega_0' = \omega_1' + \omega_2'$ and $\mathbf{k}_0 = \mathbf{k}_1 + \mathbf{k}_2$ are implied and $\beta_{ab}$ is the angle between $\mathbf{k}_a$ and $\mathbf{k}_b$: the angle of $\mathbf{k}_a$ is measured counterclockwise
with respect to $\mathbf{k}_b$. Moreover $\beta_{ab}$ is not defined at $0$ or $\pi$, and it is limited to the range of
$0<|\beta_{ab}|<\pi$. Here it is worthwhile to emphasize that the imaginary part of the frequency $\omega_a$ in the Table \ref{Ff} can be either positive or negative depending on the sign in front of it.
Moreover, six $F(asb)$'s in Table \ref{Ff} coincide with the corresponding expression Eq.(2.16) in BPE\cite{BPE99} within the wavevector inversion symmetry: in fact, no difference except the overall sign.
\begin{table}
\caption{\label{Ff}All six $F(asb)$-functions. }
\begin{ruledtabular}
\begin{tabular}{rrrrrr}
$F(asb)$ & $\omega_a$ & $\bar{\omega}_\mathbf{q}(\mathbf{k}_a : 0)$ & $\omega_b$ & $\bar{\omega}_\mathbf{q}(\mathbf{k}_b :  0)$   & $\beta_{ab}$ \\
\hline
$F(012)$ & $\omega_0$   & $(\mathbf{k}_0 : 0)$   & $\omega_2$  & $(\mathbf{k}_2 : 0)$   & $\beta_{02}$ \\
$F(201)$ & $-\omega_2$ & $(-\mathbf{k}_2 : 0)$  & $\omega_1$  & $(\mathbf{k}_1 : 0)$   & $\pi- \beta_{12}$  \\
$F(120)$ & $-\omega_1$ & $(-\mathbf{k}_1 : 0)$  & $-\omega_0$ & $(-\mathbf{k}_0 : 0)$  & $\beta_{10}$  \\
$F(021)$ & $\omega_0$  & $(\mathbf{k}_0 : 0)$    & $\omega_1$  & $(\mathbf{k}_1 : 0)$    & $-\beta_{10}$ \\
$F(102)$ & $-\omega_1$ & $(-\mathbf{k}_1 : 0)$  & $\omega_2$  & $(\mathbf{k}_2 : 0)$    & $-(\pi - \beta_{12})$  \\
$F(210)$ & $-\omega_2$ & $(-\mathbf{k}_2 : 0)$  & $-\omega_0$ & $(-\mathbf{k}_0 : 0)$  & $-\beta_{02}$  \\
\end{tabular}
\end{ruledtabular}
\end{table}

Hereafter all the wavevectors and the energies are given in the units of the Fermi wavevector
$k_F$ and the Fermi energy $\epsilon_F = \hbar \omega_F =
\hbar^2 k_F^2 /(2m_e) $ respectively, and $\hbar$ is taken to be one.

One can write the retarded qDRF of a non-interacting electron gas
$\chi(\mathbf{1} ; \mathbf{2}) \equiv \chi(\mathbf{k}_1, \omega_1 ; \mathbf{k}_2,\omega_2)$ as follows
\begin{eqnarray}
 \!\!\! \chi(\mathbf{1} ; \mathbf{2})
  & = & \frac{1}{2}
  \Big{\{} S[012] + S[021] \Big{\}}, \label{chi 2}
\end{eqnarray}
where $S[012] \equiv   F(012)   + F(201)  + F(120) $ 
and $S[021] \equiv  F(021)  + F(102) + F(210)$ are sums of three
$F(asb)$'s within the respective two cycles of the permutation.

For the convenience of calculation, one can replace the symbols of $\omega_a$ and $\omega_b$ in
$F(\mathbf{k}_a, \omega_a; \mathbf{k}_b, \omega_b)$  by the
new symbols of $\mu_a$ and  $\nu_b$ that represent an element of the set below
\begin{equation}
\Big\{\eta^\pm_a \; \Big| \; \eta^\pm_a \equiv
\frac{\pm(\omega_a' + i \delta)}{2 k_a} - \frac{k_a}{2}, \; \mbox{for} \; a = 0,1,2. \Big\}
\label{eta set}
\end{equation}
with  $\omega_a = \pm ( \omega_a' + i \delta )$ and  $k_a =|\mathbf{k}_a|$.  
This replacement comes from the relation of $\omega_a - \bar{\omega}_\mathbf{q}(\mathbf{k} _a : 0) = 2
\big( \eta_a k_a - \mathbf{q} \cdot \mathbf{k}_a \big)$.  Without $\left( \frac{1}{2} \right)^2$
factor, $F(asb) = F(\mathbf{k}_a, \mu_a; \mathbf{k}_b, \nu_b)$ can be redefined as
\begin{equation}
F(asb)  = \frac{1}{\mathcal{V}} \sum_{\mathbf{q},\sigma} n_{\mathbf{q}}
  \frac{1}{(\mu_a k_a - \mathbf{q} \cdot \mathbf{k}_a)(\nu_b k_b - \mathbf{q} \cdot
  \mathbf{k}_b)}.\quad \label{F 2}
\end{equation}
With the new definition of $F(asb)$, the qDRF can be recast into
\begin{eqnarray}
\chi(\mathbf{1} ; \mathbf{2})
   & = &  \frac{1}{8}   \Big\{  S [ 012 ] + S [ 021 ]   \Big\}. \label{chi in mu}
\end{eqnarray}

\subsection{The analytic integration of 2D $F(\mathbf{k}_a, \omega_a ; \mathbf{k}_b, \omega_b)$}

Let us get into the details of the integration of Eq.(\ref{F 2}).
In polar coordinates, it can be written as
\begin{eqnarray}
F(asb) & = & \frac{1}{2 \pi^2 k_a k_b} \int_{0}^{1} \!\! \mathrm{d} q \int_{0}^{2
\pi}\!\! \mathrm{d} \phi \nonumber \\
&& \quad \times \frac{q}{\left( \mu_a - q \cos(\phi -
\beta_{ab})\right)\; \left( \nu_b - q \cos \phi \right)}, \label{F polar}
\end{eqnarray}
where $\mu_a,\nu_b$ in $F(asb)$ can be constructed with the help of Eq.(\ref{eta set}) and the Table \ref{Ff}, and its real and imaginary parts are denoted by $\mu_a
= \mu_a' +i \mu_a'' $. The angle between $\mathbf{q}$ and $\mathbf{k}_b$ is
denoted by $\phi$.
The $\phi$-integral can be converted to a
contour integral by introducing $z = \exp(i
\phi)$ and $b = \exp(i \beta_{ab})$.  Then, one will get
\begin{eqnarray}
F(asb) & = &  \frac{- i 2b}{\pi^2 k_a k_b}
\int_{0}^{1} \!\! \mathrm{d} q \oint \mathrm{d} z \nonumber \\
&& \times \frac{z/q}{\left(
z^2 -2b \frac{\mu_a}{q} z + b^2 \right) \left(z^2 -2 \frac{\nu_b}{q} z +
1 \right)}. \label{F con}
\end{eqnarray}
The integrand has four simple poles:
\begin{subequations}
 \label{zpm}
\begin{eqnarray}
z_{\mu_a}^{\pm} & = & \frac{b}{q} \left( \mu_a \pm \sqrt{\mu_a^2 - q^2} \right), \\
z_{\nu_b}^{\pm} & = & \frac{1}{q} \left( \nu_b \pm \sqrt{\nu_b^2 - q^2} \right).
\end{eqnarray}
\end{subequations}
It is crucial to establish the analytic property of the square root complex function of $w(\mu , q) \equiv \sqrt{\mu - q^2}$ in Eq.(\ref{zpm}). The form of $w(\mu, q)$ used in Ref.\cite{R99} should be replaced by\cite{Wak,Lee08b}
\begin{equation}
\sqrt{\mu^2 - q^2}  =  \left\{
\begin{array}{ll}  \mathrm{sign}(\mu')
\sqrt{|\mu'|^2 - q^2} \qquad &\textrm{for} \;
|\mu'| > q \\ 
\mathrm{sign}(\mu'')i \sqrt{q^2-|\mu'|^2} \qquad &\textrm{for} \; |\mu'| \le q
\end{array} \right. . \label{srf}
\end{equation}
For more details of the square root complex function above, refer to the Appendix \ref{app srf}. Now one need to know which pole is inside the contour\cite{FN28}.
One can find that $z_{\mu_a}^{+}$ is always outside the contour regardless of the sign and value of
$\mu_a'$. On the other hand, $z_{\mu_a}^{-}$ is always inside the contour no matter what
value $\mu_a'$ has: a proof can be found in the appendix \ref{app poles}.  $z_{\nu_b}^{\pm}$ behaves in the same way as $z_{\mu_a}^{\pm}$ does. Considering only $z_{\mu_a}^{-}$ and $z_{\nu_b}^{-}$ inside the contour, one can use the residue theorem to reach
\begin{equation}
F (asb)
 = \frac{- i 2b}{\pi^2 k_a k_b} \int_{0}^{1} \! \mathrm{d} q \; 2 \pi i \sum residues.
\end{equation}
The residues of $z_{\mu_a}^-$ and $z_{\nu_b}^-$ can be given by $res.(z_{\mu_a}^-)$ and $res.(z_{\nu_b}^-)$ respectively as below
\begin{subequations}
\label{res F2}
\begin{eqnarray}
res.(z_{\mu_a}^-)
& = & \frac{q}{4b} \left(\frac{{\mu_a'} - \sqrt{{\mu_a'}^2 - q^2}}{-\sqrt{{\mu_a'}^2 - q^2}}\right) \frac{1}{i \sin(-\beta_{ab})} \nonumber \\
&& \times \frac{1}{  q^2 + C_\mu^+ \big({\mu_a'} - \sqrt{{\mu_a'}^2 - q^2}\big)}  \\
res.(z_{\nu_b}^-)
& = & \frac{q}{4b} \left(\frac{{\nu_b'} - \sqrt{{\nu_b'}^2 - q^2}}{-\sqrt{{\nu_b'}^2 - q^2}}\right) \frac{1}{i \sin (+ \beta_{ab}) } \nonumber \\
&& \times \frac{1}{  q^2  + C_\nu^- \big( {\nu_b'} -\sqrt{{\nu_b'}^2 - q^2}\big) },
\end{eqnarray}
\end{subequations}
where we used the fact that $w^2(\mu,q) = |\mu'|^2 - q^2$ holds true in Eq.(\ref{srf}) for both cases of $|\mu'| > q$ and $|\mu'| \le q$.  $C_{\mu}^+$ and $C_{\nu}^-$ stands for $C_\mu^+ = C(\mu_a', \nu_b', +\beta_{ab})$ and $C_\nu^- = C(\nu_b', \mu_a', -\beta_{ab})$ respectively with the definition of
\begin{equation}
C(\mu_a',\nu_b',  +\beta_{ab})  \equiv i \; \frac{\mu_a' e^{+ i \beta_{ab}} - \nu_b'}{\sin (+ \beta_{ab}) }. \label{C}
\end{equation} 

The radial $q$-integration is relatively straightforward because $C$ is merely a parameter independent of $q$. However,  depending on whether $|\mu_a'| > q$ or not, one should make the right choice in using Eq.(\ref{srf}). 

One can find that a closed-form of $F(asb)$ can be written as
\begin{eqnarray}
&& F(\mathbf{k}_a, \mu_a; \mathbf{k}_b, \nu_b) \nonumber \\
&&  =  \frac{i}{\pi \Delta_{ab}}
\left\{\mathrm{ln}\left(\frac{\varphi(\mu_a,\nu_b,+\beta_{ab})}{\varphi(\nu_b,\mu_a,-\beta_{ab})}
\right) + \ln(-e^{ + i\beta_{ab}})  \right\} , \phantom{01} \label{F function 1}
\end{eqnarray}
where $\Delta_{ab} \equiv k_a k_b \sin(\beta_{ab})$ and $\phi(\mu_a,\nu_b,+\beta_{ab})$ is 
\begin{equation}
\varphi(\mu_a,\nu_b,+\beta_{ab}) \equiv 
\mu_a' \cos(\beta_{ab})-\nu_b' - i \sin(\beta_{ab}) \sqrt{\mu_a^2-1}. \label{phi}
\end{equation}
Here it should be emphasized that $\sqrt{\mu_a^2-1}$ is the square root complex function given in the form of Eq.(\ref{srf})\cite{FN29} with $q=1$. 

Moreover, it is natural to expect Eq.(\ref{F function 1}) should be an even function of $\beta_{12}$ because there is no preferred direction to measure the angle(or chirality). To make sure, let us change the sign of $\beta_{ab}$ in Eq.(\ref{F polar}), then we get
\begin{eqnarray}
&& F(\mathbf{k}_a, \mu_a; \mathbf{k}_b, \nu_b) \nonumber \\
&&  =  \frac{i}{\pi \Delta_{ab}}
\left\{\mathrm{ln}\left(\frac{\varphi(\mu_a,\nu_b,+\beta_{ab})}{\varphi(\nu_b,\mu_a,-\beta_{ab})}
\right) -\ln(-e^{ - i\beta_{ab}})  \right\}. \phantom{01} \label{F function}
\end{eqnarray}
One would expect that Eq.(\ref{F function 1}) and Eq.(\ref{F function}) should be equivalent to each other. Indeed it looks so at a glance. 

Here it is worthwhile to emphasize that Eq.(\ref{F function 1}) and Eq.(\ref{F function}) are fully general for any arbitrary wavevector and frequency as long as $\mathbf{k}_a \ne 0$ for $a = 1, 2$ and the angles $\beta_{ab} \ne 0, \mbox{or} \; \pi$.  Moreover, there is an important issue to mention about Eq.(\ref{F function 1}) and Eq.(\ref{F function}). Once they are visualized by a computational code, they will turn out to be different from what would be expected by the analysis in the subsection \ref{ambiguity}. This discrepancy comes from the existing ambiguity in treating the phase of two logarithmic terms in a computational algorithm. The reason will be clearer later in the subsection \ref{ambiguity} and the more feasible form will be given as the single logarithmic form in Eq.(\ref{F function single}). 

Before we go over the static features of the 2D qDRF, it is worthwhile to mention how $F(asb)$ behaves in the high-frequency limit or in the static short-wavelength limit.  Assuming $|\mu_a'| > 1$ and $|\nu_b'| >1$, $F(\mathbf{k}_a, \mu_a; \mathbf{k}_b, \nu_b)$ approaches zero as $\mu_a'$ or $\nu_b'$ goes to the infinity, even it does so as $\sin \beta_{12}$ in $\Delta$ goes to zero because the curly bracket in Eq.(\ref{F function 1}, \ref{F function}, \ref{F function single}) approaches zero whereas the static 2D qDRF obtained by Zhang\cite{Z91} can diverge as $\sin \beta_{12}$ goes to zero.

\section{\label{ch static}The static $\chi(\mathbf{k}_1; \mathbf{k}_2)$ in 2D}

As for the exact static case($\omega_1 = \omega_2 = 0$), $S[012]$ and $S[021]$ are complex conjugates to each other due to the fact of $F^*(asb) = F(bsa)$:\cite{FN30}
\begin{subequations}
\label{sym F3}
\begin{eqnarray}
F^*(012)  & = & F(210)  \\
F^*(201) & = &   F(102)  \\
F^*(120) & = &  F(021). 
\end{eqnarray}
\end{subequations}
Therefore the static $\chi(\mathbf{k}_1 ; \mathbf{k}_2) \equiv \chi (\mathbf{k}_1, 0; \mathbf{k}_2, 0)$ can
be written as
\begin{eqnarray}
\chi(\mathbf{k}_1; \mathbf{k}_2)
   & = & \frac{1}{4} S'[012], \label{chi s}
\end{eqnarray}
where $S'$ is the real part of $S = S' + i S''$.
The notation of Eq.(\ref{chi s}) means the imaginary part of the static $\chi(\mathbf{k}_1; \mathbf{k}_2)$ is simply zero. 

\begin{figure}
\centering
    \includegraphics[width=.5\textwidth]{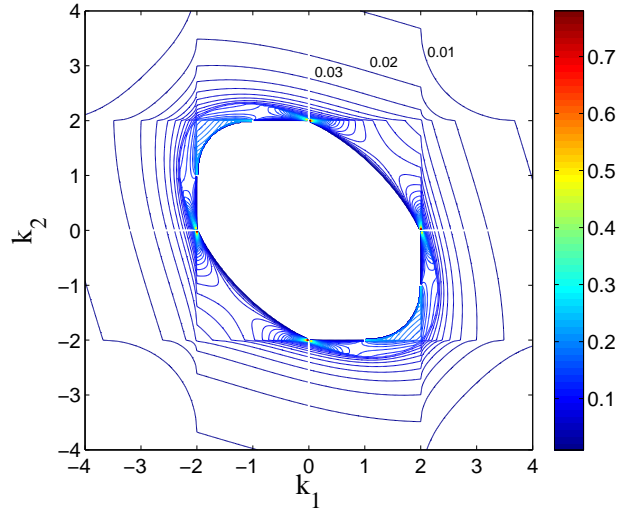}
    \vspace{-0.5cm}
    \caption{(Color online) The contour plot of the static $\chi'(\mathbf{k}_1; \mathbf{k}_2)$. The plot points are separated by $\Delta k_1 = \Delta k_2 = 0.02$. The difference between consecutive contours is $0.01$. The first quadrant stands for $\chi'(\mathbf{k}_1, \mathbf{k}_2)$ with the angle of $\beta_{12}= 60^\circ$. The second quadrant stands for $\chi'(-\mathbf{k}_1; \mathbf{k}_2)$ with $\beta_{12} =120^\circ$.  $\chi(\mathbf{k}_1; \mathbf{k}_2)$ is not defined at $\mathbf{k}_1 = 0$ nor at $\mathbf{k}_2 = 0$. }
  \label{contour60}
\end{figure}

For a given angle of $\beta_{12}$, the real part $\chi'(\mathbf{k}_1; \mathbf{k}_2)$ can be plotted in the Cartesian coordinates $(k_1,k_2)$. Let us take a look at a contour plot shown in Fig.(\ref{contour60}). The first quadrant stands for $\chi'(\mathbf{k}_1; \mathbf{k}_2)$ with the angle of $\beta_{12}= 60^\circ$.
The negative axis of $k_1$($k_2$) represents the inversion of the wavevector of $\mathbf{k}_1$($\mathbf{k}_2$), so that the second quadrant stands for $\chi'(-\mathbf{k}_1; \mathbf{k}_2)$ with $\beta_{12}  =120^\circ$(between $-\mathbf{k}_1$ and $\mathbf{k}_2$). The two
bottom quadrants are the reflection of the upper ones with respect to the origin, which shows merely the inversion symmetry $\chi(\mathbf{k}_1; \mathbf{k}_2) = \chi(-\mathbf{k}_1; -\mathbf{k}_2)$. 

A number of interesting static features can be observed in Fig.(\ref{contour60}) such as \textit{vanishing, discontinuity}, and \textit{peaks}. In the following subsections, we will take a look at these three features one by one. In addition, an ambiguity will be reported in treating the phase of $F(asb)$ in Eq.(\ref{F function}), which will turn out to cause the overall sign ambiguity of the 2D qDRF.

\subsection{\label{vanishing}Vanishing}

The first easily noticeable feature is the quite broad central
region in Fig.(\ref{contour60}) where $\chi'(\mathbf{k}_1; \mathbf{k}_2)$ vanishes. This feature of vanishing persists in the static $\chi'(\mathbf{k}_1; \mathbf{k}_2)$ throughout the angle $\beta_{12}$ from $0$ to $\pi$.  

\begin{figure*}
\centering     
  \begin{subfigure}[b]{.33\linewidth}
    \centering
    \includegraphics[width=.99\textwidth]{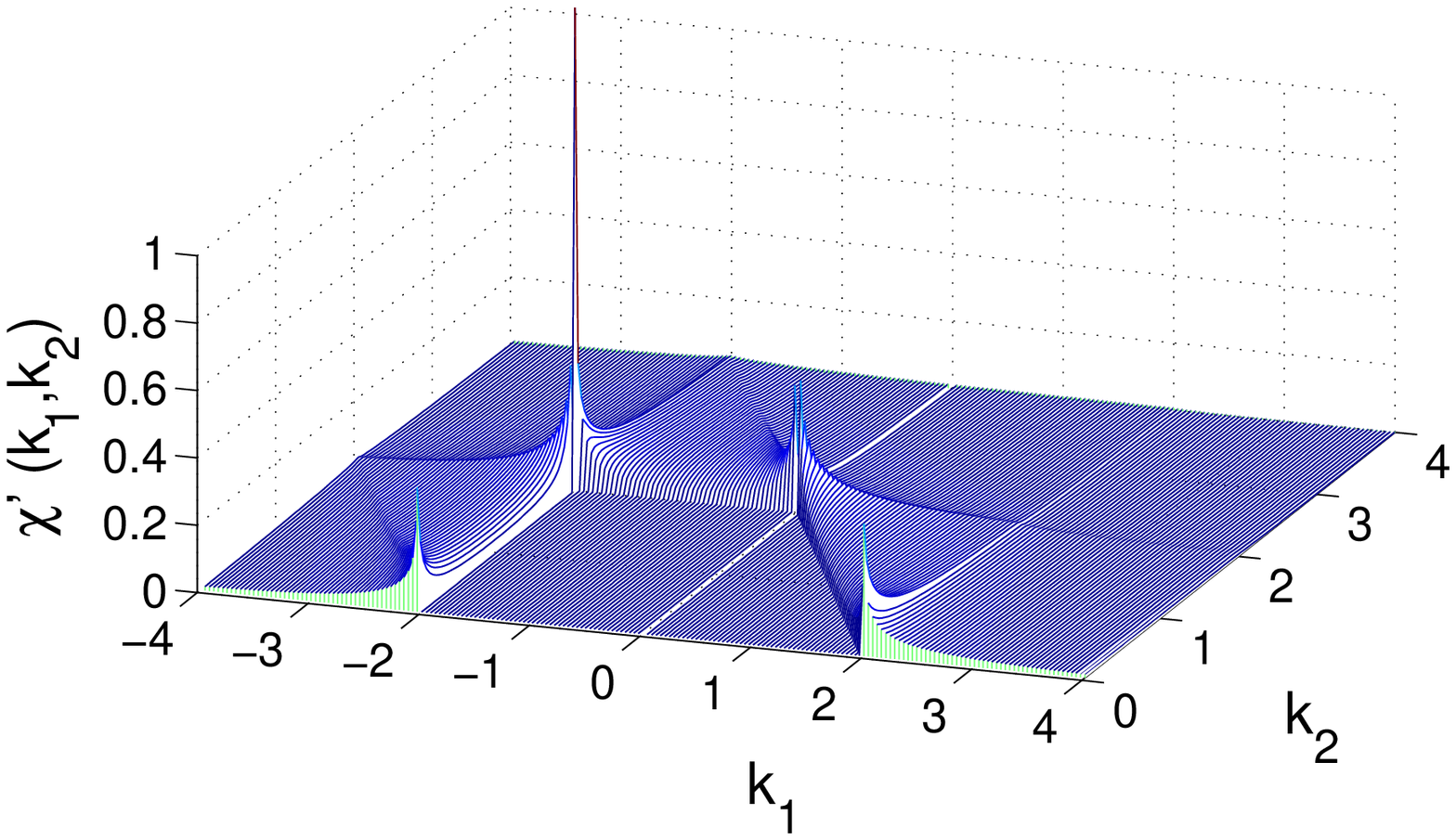} \vspace{-0.5cm}
    \subcaption{$\beta_{12} = 5^\circ$.}\label{fig:3da}
  \end{subfigure}%
  \begin{subfigure}[b]{.33\linewidth}
    \centering
    \includegraphics[width=.99\textwidth]{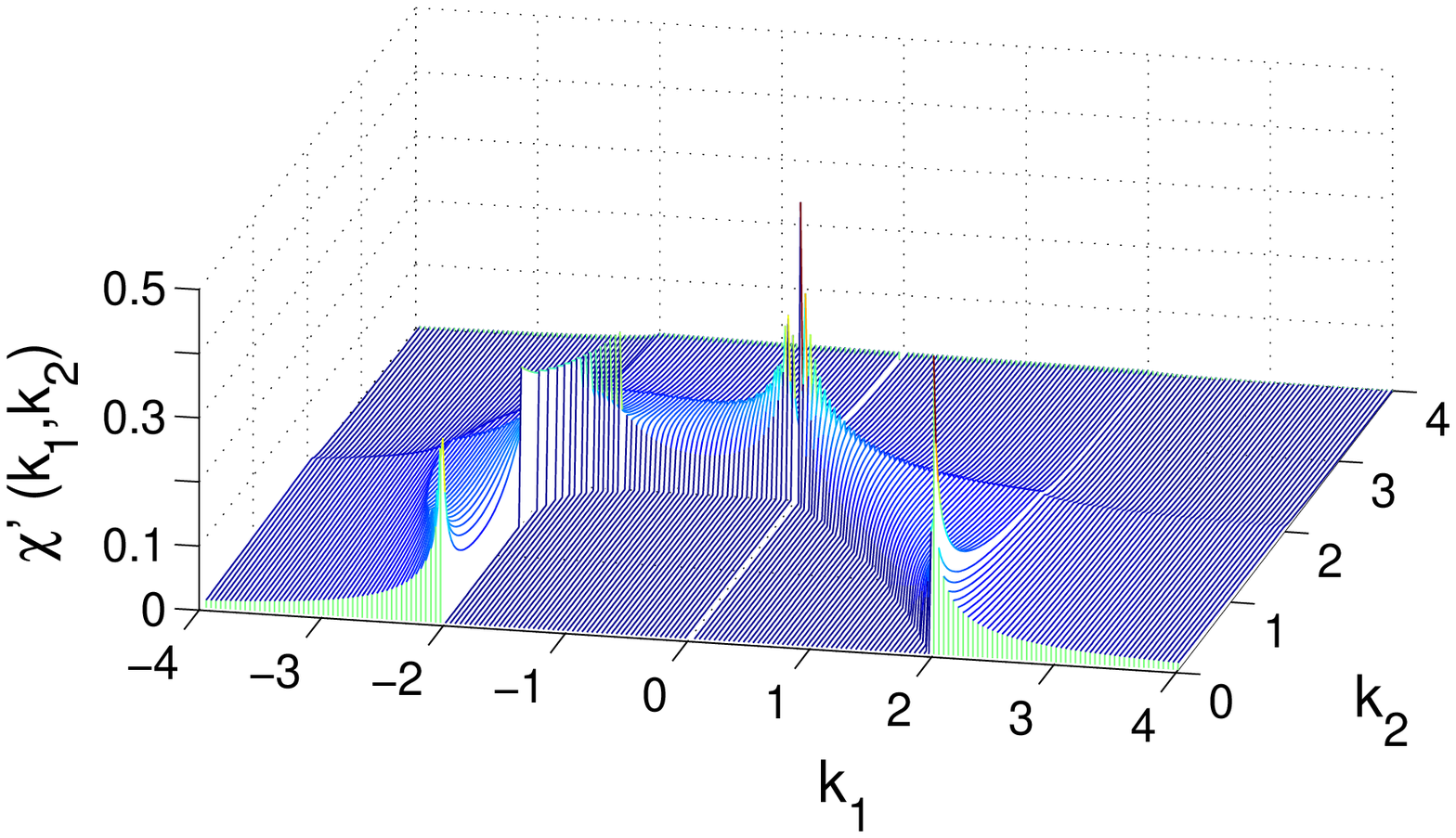} \vspace{-0.5cm}
    \subcaption{$\beta_{12} = 45^\circ$.}\label{fig:3db}
  \end{subfigure}%
  \begin{subfigure}[b]{.33\linewidth}
    \centering
    \includegraphics[width=.99\textwidth]{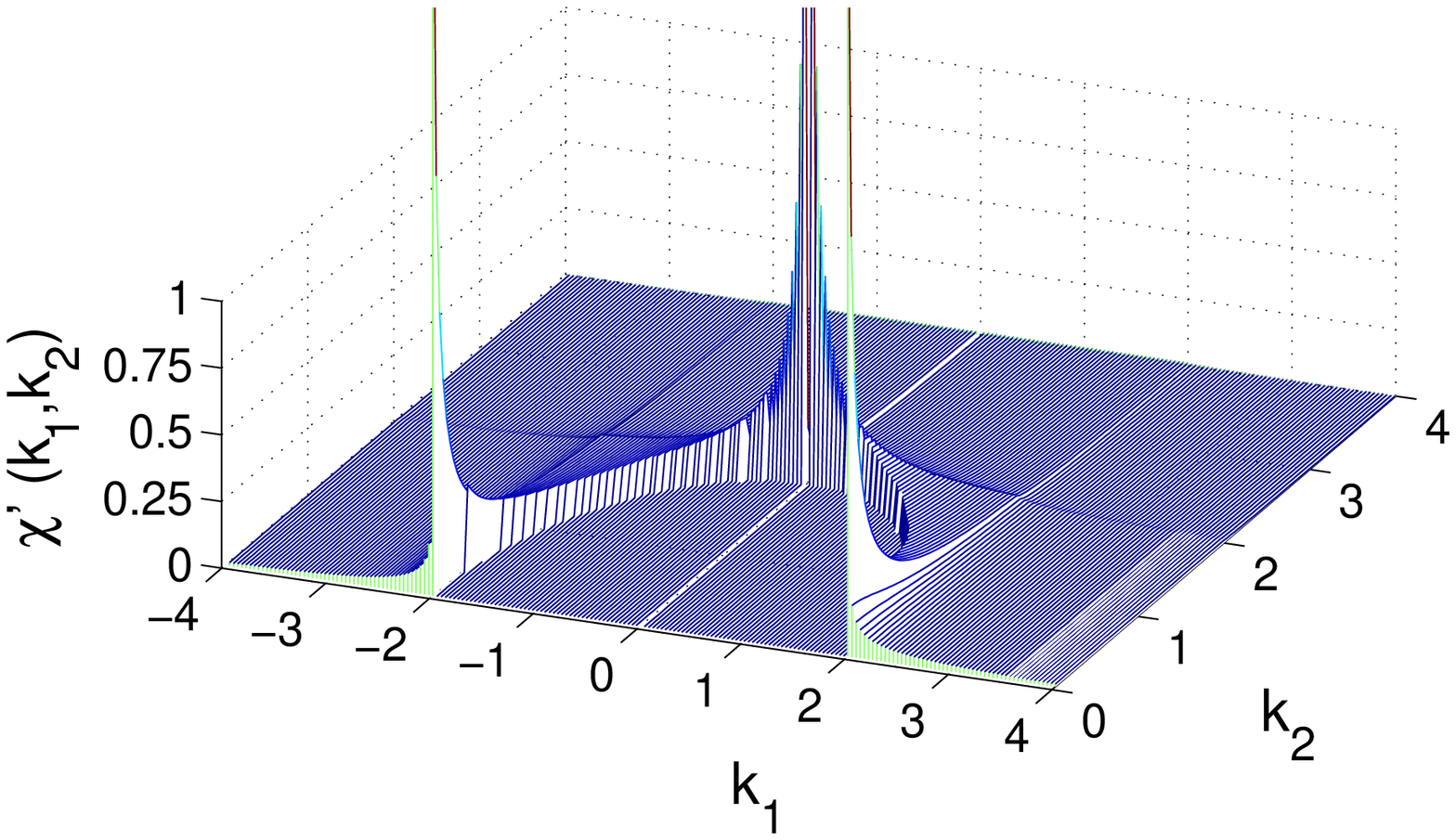} \vspace{-0.5cm}
    \subcaption{$\beta_{12} = 90^\circ$.}\label{fig:3dc}
  \end{subfigure}\\%
  \begin{subfigure}[b]{.165\linewidth}
    \centering
    \includegraphics[width=.99\textwidth]{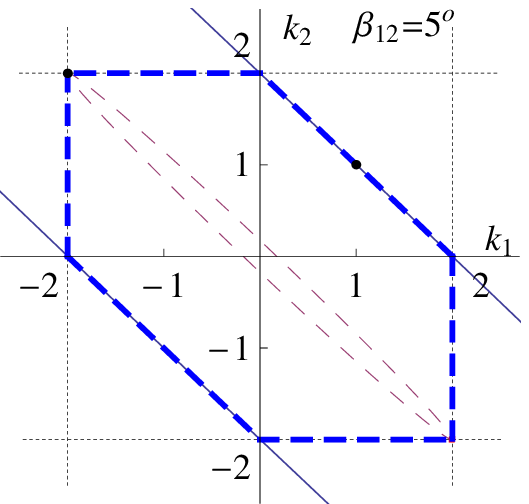}
    \subcaption{$\beta_{12} = 5^\circ$.}\label{fig:a5}
  \end{subfigure}%
  \begin{subfigure}[b]{.165\linewidth}
    \centering
    \includegraphics[width=.99\textwidth]{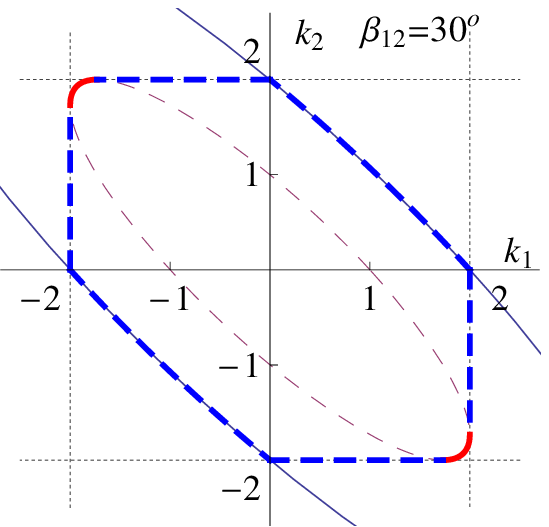}
    \subcaption{$\beta_{12} = 30^\circ$.}\label{fig:a30}
  \end{subfigure}%
  \begin{subfigure}[b]{.165\linewidth}
    \centering
    \includegraphics[width=.99\textwidth]{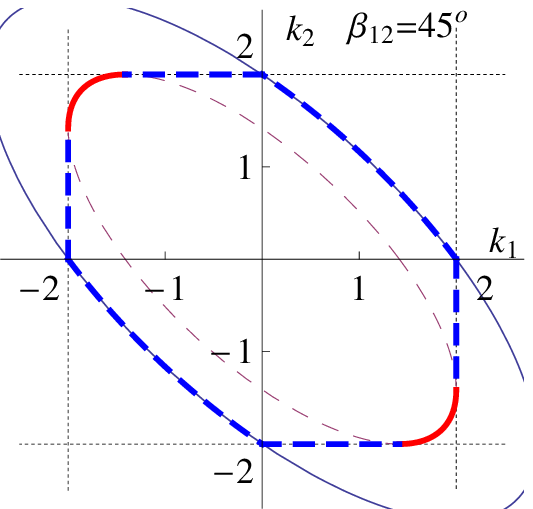}
    \subcaption{$\beta_{12} = 45^\circ$.}\label{fig:a45}
  \end{subfigure}%
  \begin{subfigure}[b]{.165\linewidth}
    \centering
    \includegraphics[width=.99\textwidth]{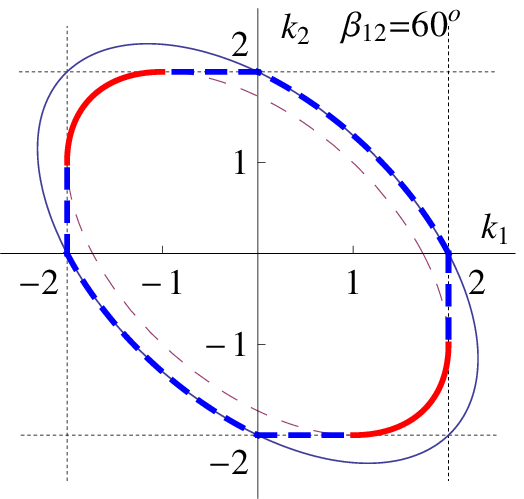}
    \subcaption{$\beta_{12} = 60^\circ$.}\label{fig:a60}
  \end{subfigure}%
  \begin{subfigure}[b]{.165\linewidth}
    \centering
    \includegraphics[width=.99\textwidth]{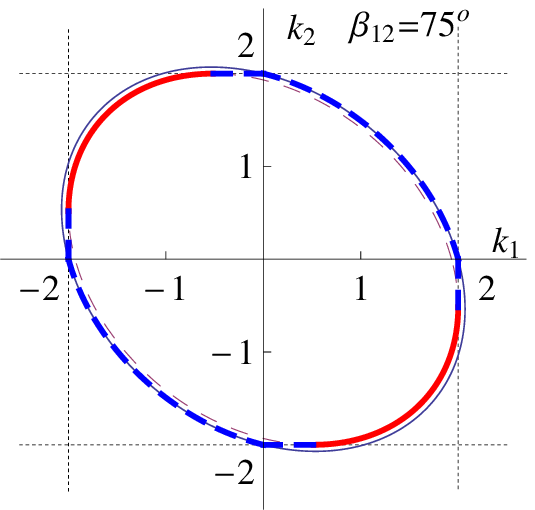}
    \subcaption{$\beta_{12} = 75^\circ$.}\label{fig:a75}
  \end{subfigure}%
  \begin{subfigure}[b]{.165\linewidth}
    \centering
    \includegraphics[width=.99\textwidth]{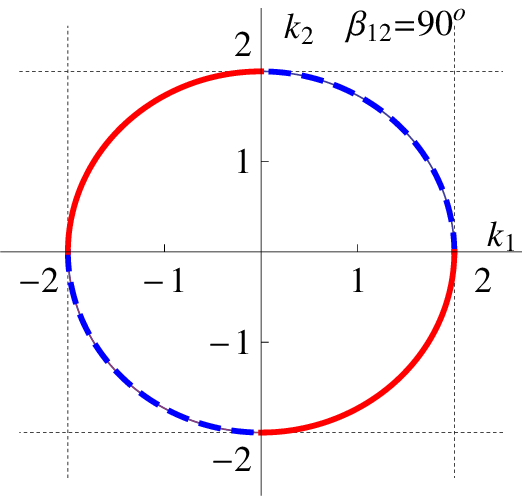}
    \subcaption{$\beta_{12} = 90^\circ$.}\label{fig:a90}
  \end{subfigure}\vspace{-0.2cm} 
   \caption{(Color online)  (a)-(c): The 3D plots of the static $\chi'(\mathbf{k}_1; \mathbf{k}_2)$ at three different angles: $\beta_{12} = 5^\circ, 45^\circ, 90^\circ$. (d)-(i): The angle dependence of the boundary of the vanishing region. If $\beta_{ab} > 90^\circ$, the left and right sides would be interchanged.}
  \label{fig:3d}
\end{figure*}

The 3D plots of 
Figs.(\ref{fig:3da})-(\ref{fig:3dc}) portray $\chi'(\mathbf{k}_1; \mathbf{k}_2)$ at three different angles $\beta_{12} = 5^\circ, 45^\circ, 90^\circ$. Figs.(\ref{fig:a5})-(\ref{fig:a90}) show how the shape of the vanishing region changes as $\beta_{12}$ varies from $5^\circ$ to $90^\circ$. The boundary of the vanishing region is depicted by the thick dashed(blue) and thick solid(red) lines. 

There is a simple way to construct the vanishing region. Firstly, focus on the $4 \times 4$ square in $(k_1, k_2)$-space, and cut off a part beyond the outer ellipse. Final touch is just cutting off the remaining part beyond the inner ellipse only at the corner of the second quadrant. 

This vanishing region can be described by a number of mathematical conditions.
First, it is just the common region satisfying the following three conditions:
\begin{subequations}
 \label{k con}
\begin{eqnarray}
 |\eta_0'|  = |k_0/2|  & <  & 1, \label{k con 0}\\
 |\eta_1'|  = |k_1/2|  & <  & 1, \label{k con 1}\\
 |\eta_2'|  = |k_2/2|  & <  & 1. \label{k con 2}
\end{eqnarray}
\end{subequations}
Eq.(\ref{k con 1}) and Eq.(\ref{k con 2}) construct the $4\times4$ square whereas Eq.(\ref{k con 0}) cuts off a part beyond the outer ellipse($|\eta_0'|=1$). Final touch is the exclusion of the common region described by
\begin{equation}
\varphi^+ \varphi^-_{asb}  \ge 0 \quad \textrm{and} \quad \ell^+ \ell^- _{asb} < 0, \label{ex con}
\end{equation}
where $\varphi^+ \varphi^-_{asb} = 0$ is the inner ellipse whereas $\ell^+ \ell^- _{asb} < 0$ is chosen to select the corner in the second quadrant.
The $\varphi^\pm_{asb}$ is defined by $\varphi^\pm_{asb} \equiv \varphi(\eta_a', \eta_b', \pm \beta_{ab})$ whereas $\ell^+_{asb}$ and $\ell^-_{asb}$ are defined by $\ell^+_{asb} \equiv \eta_a' \cos(+ \beta_{ab}) - \eta_b'$ and $\ell^-_{asb} \equiv \eta_b' \cos(- \beta_{ab}) - \eta_a'$ respectively. One should construct $\eta_a'$ and $\eta_b'$ from the Table \ref{Ff} to find the specific expressions of $\varphi^+\varphi^-_{asb}$ and $\ell^+ \ell^-_{asb}$ for each $F(asb)$.  One will find that three $\varphi^+\varphi^-_{asb}$'s in $S[012]$ are identical\cite{FN31}, so that they can be denoted by the single notation  $\varphi^+\varphi^-$.

\begin{figure}
\centering
  \includegraphics[height=0.25\textwidth]{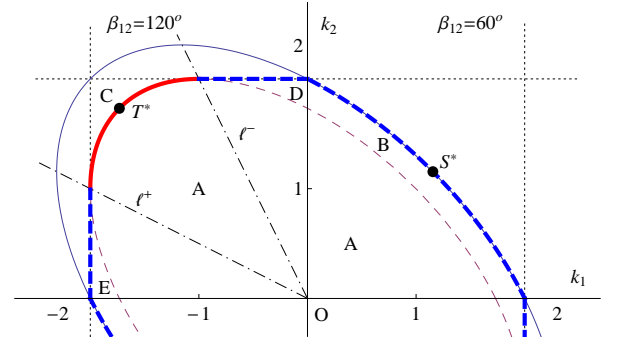}
  \caption{(Color online) Vanishing region of Fig.(\ref{contour60}) determined by the conditions of Eq.(\ref{k con}) and Eq.(\ref{ex con}). $|k_0/2| =1$ corresponds to the outer ellipse and
    $\varphi^+\varphi^- =0$ stands for the inner dashed ellipse. }\label{Qvanish}
\end{figure}

For clarity, let us take a look at  Fig.(\ref{Qvanish}), which highlights the vanishing boundary in the upper quadrants of Fig.(\ref{contour60}). The $4\times4$ square is denoted by the dotted lines($|k_1 / 2| = 1, \;|k_2 / 2| = 1$). The outer ellipse($|k_0/2|  = 1$) and the inner ellipse($\varphi^+\varphi^-  = 0$) are denoted by the solid line and the dashed line respectively. In between the dot-dashed lines($\ell^+ ,  \ell^-$) resides the corner of the second quadrant. Then, the exclusion condition in Eq.(\ref{ex con}) removes the region C from the vanishing region, so that the vanishing region is enclosed with only the thick lines in Fig.(\ref{Qvanish}):  refer to the subsection \ref{discon} for the mathematical mechanism of the vanishing.

Let us go back to Fig.(\ref{Qvanish}) and take a look at the dots: $S^*$ and $T^*$. They are symmetric points where $k_1 = k_2$.  What is the distance of $OS^*$($OT^*$)? At first sight, it seems difficult to answer the question. However, once we adopt the contra-variant oblique coordinates, the answer is easy to find: refer to Appendix \ref{ssB} for more details. In order to present the answer, let us denote the components of $S^*$($T^*$) with $(k_1^*, k_1^*)$, the distance of $OS^*$($OT^*$) is simply $\sqrt{2} k_1^*$ where\cite{FN32}
\begin{equation}
  k_1^* (\beta_{12}) = \left\{ 
\begin{array}{ll} 
\sqrt{2}/\sqrt{1 + \cos\beta_{12}}             \quad & \mathrm{for} \;\; \beta_{12} \le \pi/2 \\ 
\sqrt{2} \sin \beta_{12}/\sqrt{1 +\cos \beta_{12}} \;\;  \quad  & \mathrm{for} \;\; \beta_{12} > \pi/2
\end{array} \right.  \!\! . \label{k1 beta}
\end{equation}

\begin{figure}
\centering
  \includegraphics[height=0.35\textwidth]{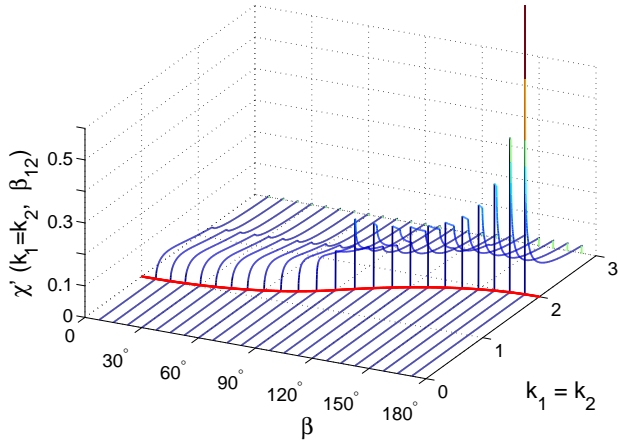}
  \caption{(Color online) The static 2D $\chi'(\mathbf{k}_1; \mathbf{k}_2)$ when $k_1 = k_2$. The angle $\beta_{12}$ varies  from $0$ to $\pi$ by $\pi/24$ increment. }
  \label{bk}
\end{figure}

What does $\chi'(\mathbf{k}_1; \mathbf{k}_2)$ look like along the direction of $OS^*$($OT^*$) in Fig.(\ref{Qvanish})?  Along this symmetric line, we know $k_1 = k_2$ so that it depends only on $k_1$ at various angle $\beta_{12}$. Let us take a look at Fig.(\ref{bk}), which shows how $\chi'(\mathbf{k}_1; \mathbf{k}_2)$ looks as $\beta_{12}$ varies from $0$ to $\pi$. We can clearly see the vanishing line up to $S^*$ for $\beta_{12} < \pi/2$ and otherwise up to $T^*$. Interestingly the discontinuous jump takes place only for $\beta_{12} > \pi/2$. The thick solid(red) curve follows $k_1^*(\beta_{12})$ in Eq.(\ref{k1 beta}) showing a perfect match with the end points($S^*, T^*$) of the vanishing lines. 

Quite remarkably, the four conditions of Eq.(\ref{k con}) and Eq.(\ref{ex con}) are tantamount to the \textit{triangle rule}, which states that the vanishing occurs only if the wavevector triangle(composed of $\mathbf{k}_1, \mathbf{k}_2, -\mathbf{k}_0$) fits inside the Fermi circle: refer to a simple geometrical proof in Appendix \ref{app triangle}. Therefore, the vanishing boundary is closely related to the existence of the sharp Fermi surface of a 2DEG. This feature of the \textit{triangle rule} was correctly pointed out by Rommel and Kalman\cite{R99}.

As a matter of fact, this vanishing feature in a 2DEG is not entirely unexpected but required by the quadratic compressibility sum rule\cite{GKD75, RK96, KR00a}
\begin{equation}
\! \! \lim_{\mathbf{k}_1, \mathbf{k}_2 \rightarrow 0} \! \! \chi' (\mathbf{k}_1; \mathbf{k}_2) = \frac{n}{2} \frac{(\partial p/\partial n)_T - n (\partial^2 p / \partial n^2 )_T}{(\partial p / \partial n)^3_T} = 0,
\label{q comp}
\end{equation}
which implies the vanishing in the static long-wavelength limit because the pressure $p$ of a 2DEG is proportional to $ n^2$  at zero temperature. This vanishing in the static long-wavelength limit can be verified by our closed-form in Eq.(\ref{F function}) in a direct manner: a proof can be found in Appendix \ref{app long w}. On the other hand, the static 2D qDRF obtained by Zhang\cite{Z91} does not show the vanishing. Motivated by this discrepancy, one might wonder why a \textit{quantum} 2DEG should surely satisfy the \textit{classical} quadratic compressibility sum rule of a 2DEG. In fact, it was revealed that the plasma frequency in a relativistic quantum 2DEG does not show the same density dependence as that of a non-relativistic quantum 2DEG in the classical limit\cite{HDS07,DSH09}. Nevertheless, a non-relativistic quantum 2DEG may differ from its relativistic counterpart.  According to Barton\cite{B83}, the scattering amplitude of a non-relativistic quantum 2DEG becomes identical to the classical one in the static long-wavelength limit, so that the classical quadratic compressibility sum rule, Eq.(\ref{q comp}), is expected to be valid for a quantum 2DEG either.

Moreover, it is noteworthy that $\chi'(\mathbf{k}_1; \mathbf{k}_2)$ satisfies the collinear limit expression Eq.(3.25) in PRE\cite{PRE95} found for a 3DEG. In the parallel collinear limit$(\mathbf{k}_1 \parallel \mathbf{k}_2 )$ as $\beta_{12}$ goes to $0$, the 2D qDRF of $\chi'(\mathbf{k}_1; \mathbf{k}_2)$ can be rewritten in terms of the 2D linear DRFs as follows
\begin{equation}
\chi'  (\mathbf{k}_1; \mathbf{k}_2) = + \frac{1}{k_0 k_1 k_2} 
\Big\{ \! k_0  \chi' (\mathbf{k}_0 \! ) - k_1 \chi' (\mathbf{k}_1 \! ) - k_2 \chi' (\mathbf{k}_2 \!) \!
\Big\}. \label{collinear}
\end{equation}
Here $\mathbf{k}_0 = \mathbf{k}_1 + \mathbf{k}_2$ is the largest wavevector. If $\mathbf{k}_1$ and $\mathbf{k}_2$ are anti-parallel $(\mathbf{k}_1 \parallel -\mathbf{k}_2 )$, the subscripts of the wavevectors should be reordered in such a way that the largest wavevector comes first in the curly bracket. This collinear-limit expression can be derived directly from Eq.(\ref{F function single}): a proof can be found in Appendix \ref{app col}. In comparison with Eq.(3.25) in PRE\cite{PRE95}, Eq.(\ref{collinear}) is two times smaller and its sign is opposite. The Eq.(\ref{collinear}) also shows the vanishing in the static long-wavelength limit and, to a certain extent, even at finite $(k_1,k_2)$-values too. This vanishing feature in the collinear limits also validates the classical quadratic compressibility sum rule even for a quantum 2DEG.  In addition, Eq.(\ref{collinear}) implies $\chi' (\mathbf{k}_1; \mathbf{k}_2)$ does not change its sign remaining non-negative in the entire $(k_1, k_2)$-space\cite{FN35} or positive in the non-vanishing region.

By looking at Fig.(\ref{bk}) when $\beta_{12}$ is close to $0$ or $\pi$, one can guess what $\chi' (\mathbf{k}_1; \mathbf{k}_2)$  looks like in the collinear limit.  As $\beta_{12}$ gets closer to 0 in the parallel collinear limit, the non-vanishing part of $\chi' (\mathbf{k}_1; \mathbf{k}_2)$ does not diminish to zero but remain finite for $k_1 > 1$ while it develops the peak-like behavior around $k_1 = 2$  as $\beta_{12}$ gets closer to $\pi$ in the anti-parallel collinear limit.

\subsection{\label{discon}Discontinuity}

The second static feature of the 2D qDRF is the discontinuity when $\beta_{12} > \pi/2$. It takes place 
across the vanishing boundary of $\varphi^+ \varphi^- = 0$ in Fig.(\ref{Qvanish}), 
where the discontinuous boundary is denoted as the thick solid
line in the second quadrant. For clarity, take a look at the discontinuous jump in the second quadrant of Fig.(\ref{fig:3db}) as well as the region in Fig.(\ref{bk}) where $\beta_{ab} > 90^\circ$.
This non-analytic feature is due to the abrupt phase-change of the logarithmic terms of $F(asb)$ in Eq.(\ref{F function}). 

To look into the mechanism of the discontinuity, let us analyze the phases of $F(asb)$ in Eq.(\ref{F function}) focusing on only the common region in which $|\eta_a'|<1$ for $a = 0,1,2$. $F(asb)$ in Eq.(\ref{F function}) has two logarithmic terms in the curly bracket: simply call them $\ln \big( \varphi^+_{ab} / \varphi^-_{ba} \big) -\ln(-e^{-i\beta_{ab}})$. The phase of the former is just $0$ or $\pi$ because any $\varphi^{+}_{ab}$($\varphi^{-}_{ba}$) becomes a real function inside the region of our interest: see Eq.(\ref{phi}). On the other hand, the phase of the latter is simply $-(\pi - \beta_{ab})$. Therefore, the phase of the sum becomes just $(0 \; or \; \pi) - (\pi - \beta_{ab})$.

Let us introduce the new symbol of $\vartheta(asb) \equiv \vartheta_1(asb) + \vartheta_2(asb)$ with $\vartheta_1(asb) \equiv (0 \; or \; \pi)$ and $\vartheta_2(asb) \equiv - (\pi - \beta_{ab})$.  Then, the real part of $S'[012]$ can be written as
\begin{equation}
S'[012] =  -\frac{1}{\pi \Delta} \vartheta[012],
\end{equation}
where $\Delta \equiv \Delta_{02} = \Delta_{21} = \Delta_{10}$ is the area of the wavevector triangle and $\vartheta[012] \equiv \vartheta(012) + \vartheta(201) + \vartheta(120)$.  As given in Table \ref{phase}, the phase $\vartheta(asb)$ in $S[012]$ varies depending on the region from A to E in Fig.(\ref{Qvanish}). 
\begin{table}
\caption{\label{phase} Phase $\vartheta(asb)$ for each $F(asb)$-function. Refer to the region of A,B,C,D and E in Fig.(\ref{Qvanish}).}
\begin{center}
\begin{ruledtabular}
\begin{tabular}{lrrrr}
$\vartheta(asb)$ & A,B & C & D & E \\
\hline
$\vartheta(012)$ & $ \beta_{02}$ & $- \pi +\beta_{02} $ & $\beta_{02}$ & $-\pi + \beta_{02}$ \\
$\vartheta(201)$ &  $-\beta_{12}$ & $ -\beta_{12}$ & $\pi -\beta_{12}$ & $\pi - \beta_{12}$ \\
$\vartheta(120)$ & $\beta_{10}$ & $- \pi + \beta_{10}$ & $-\pi + \beta_{10}$ & $\beta_{10}$ \\
\hline
$\vartheta[012]\footnote{$\vartheta[012] \equiv \vartheta(012) + \vartheta(201) + \vartheta(120)$}$ &  0 & $-2\pi$  & 0 & 0 \\
\end{tabular}
\end{ruledtabular}
\end{center}
\end{table}
The phases of $\vartheta(asb)$ in $S[012]$ cancel each other out only in the vanishing region once they are summed up\cite{FN36}. However, they don't fully cancel each other out in the region of C ending up with $\vartheta[012]= -2 \pi$. Therefore,  across the vanishing boundary from A to C, the phase of $\vartheta[012]$ discontinuously changes from $0$ to $-2\pi$  leading the value of $S'[012]$ to jump from 0 to $2/\Delta$.

\subsection{Peaks}

The third static feature of the 2D qDRF is the strong peaks at some points along the vanishing boundary, where the ellipse($|{\eta_0'}^\pm| =1$ or $\varphi^+ \varphi^- = 1$) meets the condition of $k_a= 2 k_F$ for $a=1,2$ that is reminiscent of Kohn anomaly\cite{Kohn59,KL65}: for example, $(k_1, k_2) =
(2,0)$, $(0,2)$, $(-2, 2\cos\beta_{12})$ or $(-2\cos\beta_{12}, 2)$ with $\beta_{12} = \pi/3$ in Fig.(\ref{Qvanish}). Even though the values of these peaks are not visualized
correctly in Figs.(\ref{fig:3da})-(\ref{fig:3dc}), one can see a certain
tendency of the peak-like singular behavior.  

The angle $\beta_{12} = \pi/2$ is
special in a sense that the inner and the
outer ellipses become identical to a circle with the radius of $2$ as shown in Fig.(\ref{fig:3dc}) and Fig.(\ref{fig:a90}).  
For $\beta_{12} = \pi/2$, the $k_0 = 2k_F$ condition is satisfied all the way around 
the vanishing boundary. 

In fact, the conditions of the strong peaks coincide with the condition of $\Delta_{ab} \equiv k_a k_b \sin \beta_{ab} = 0$ that makes $F(asb)$-function diverge. In other words, on the peak condition of $\Delta_{ab} = 0$, a particle-hole pair excitation and recombination takes place between the opposite sides of the Fermi surface. Such a particle-hole pair bubble corresponds to the one-loop triangle diagram with one of its sides pinched.

As for the points at the edge of the discontinuous boundary, we can expect peaks from $\Delta_{ab} = 0$ only in the anti-parallel collinear limit as $\beta_{ab}$ goes to $\pi$. Remind that Eq.(\ref{F function}) is not well-defined at $\beta_{ab} = \pi$, however its collinear-limit expression in Eq.(\ref{collinear}) shows a strong peak-like behavior. 

What about the parallel collinear limit? As mentioned in the subsection \ref{vanishing}, $\chi'(\mathbf{k}_1 ; \mathbf{k}_2)$ in Fig.(\ref{bk}) remains finite for $k_1 > 1$  as $\beta_{12}$ goes to 0. This fact may lead one to conclude that there is no peak in the parallel collinear limit. However, that is misleading because the plot of $\chi'(\mathbf{k}_1 ; \mathbf{k}_2)$ in Fig.(\ref{bk}) is just the special case of $k_1 = k_2$. In fact, there are another two peaks at $(k_1, k_2)_O = (2, 0), (0,2)$ that come close to the point $S^*$ from the both sides of it along the vanishing boundary: refer to the contra-variant oblique coordinates $(k_1, k_2)_O$ in Appendix \ref{ssB} and decease the angle $\beta_{12}$ to see how the vanishing region and the peak's points change.

\subsection{\label{ambiguity}Ambiguity}

In the foregoing subsection of discontinuity, we added up $\vartheta_1(asb)$ and $\vartheta_2(asb)$ to get $\vartheta(asb)$ before we summed up three phases of $\vartheta(asb)$ to get $\vartheta[012]$.  This is actually the way how we have obtained the plot of $\chi'$ for a visualization.

There is another distinct way to sum up the phase of $\vartheta(asb)$ in $S[012]$. As for this second case, pick up only $\vartheta_2(asb)$ from $\vartheta(asb)$ to get $\vartheta_2[012]$ and also get $\vartheta_1[012]$ in the same way before we sum up $\vartheta_1[012]$ and $\vartheta_1[012]$ to get $\vartheta[012]$. 

One may expect that the second way to sum up the phases is equivalent to the first case. However, depending on which way one sums up the phase $\vartheta(asb)$, the sign and value of $S[012]$ will turn out to be different: refer to Appendix \ref{app am} for more details. 

Unfortunately this ambiguity brings on the overall sign ambiguity accompanied by the value of $\vartheta[012]$ itself changed.
From the mathematical point of view, this ambiguity seems related to the well-known fact that $\ln e^{z}$ is not always equal to $z$. Indeed, it is not appropriate to simply set $\ln \left(-e^{-i \beta_{ab}} \right) = - (\pi - \beta_{ab})$ by treating it as an independent term for the purpose of the numerical calculation(or a visualization). 
To avoid this ambiguity, one should change the expression of $F(asb)$ in Eq.(\ref{F function}) into a single logarithmic term as follows\cite{FN39}
\begin{eqnarray}
&& F(\mathbf{k}_a, \mu_a; \mathbf{k}_b, \nu_b) \nonumber \\
&&  =  \frac{i}{\pi \Delta_{ab}}
   \mathrm{ln} \left\{  -e^{ i\beta_{ab}}\left(  \frac{\varphi(\mu_a,\nu_b,+\beta_{ab})}{\varphi(\nu_b,\mu_a,-\beta_{ab})} \right) \right\}, \label{F function single}
\end{eqnarray}
even if there still remains an ambiguity in treating the phase, especially when one try to extract the real part for a numerical calculation. In fact it is the Eq.(\ref{F function single}) that has been used to draw the plot of $\chi'$ for a visualization\cite{FN40}. 
This single logarithmic form in Eq.(\ref{F function single}) guarantees that $F(asb)$ does not diverge in the static short-wavelength limit or in the high-frequency limit, and naturally the 2D qDRF $\chi(\mathbf{k}_1, \omega_1 ; \mathbf{k}_2,\omega_2)$ does not so.

\section{\label{conc}Conclusion}

A closed-form expression of the retarded quadratic density response function(qDRF) of a non-interacting 2DEG has been obtained, which has been expressed in terms of a wavevector- and frequency-dependent non-analytic complex function. An emphasis has been placed on a number of exact static features of the 2D qDRF such as \textit{vanishing}, \textit{discontinuity}, \textit{peaks}, and \textit{ambiguity} that have not been seen in any other dimension.

Firstly, the static $\chi(\mathbf{k}_1; \mathbf{k}_2)$ vanishes when it satisfies \textit{triangle rule}: it vanishes whenever the wavevector-triangle fits inside the Fermi circle irrespective of the
angle $\beta_{12}$.  We specified the vanishing region and explained its
vanishing mechanism mathematically. However, the physical meaning and its origin of the vanishing is still a puzzle.

Secondly, when $\beta_{12} > \pi/2$, the abrupt response appears discontinuously across the vanishing boundary of $\varphi^+ \varphi^- = 0$. This non-analytic feature of the discontinuity is related to the phase jump of the logarithmic function in $F(asb)$-function.

Thirdly, some characteristic peaks can be seen along the vanishing boundary only in case the ellipse($|{\eta_0'}^\pm| =1$ or $\varphi^+ \varphi^- = 1$) satisfies the $k_a = 2k_F$ condition for $a =1,2$. Especially for $\beta_{12} = \pi/2$, the $k_0 = 2k_F$ condition  is always satisfied all the way through the vanishing boundary. 

Finally, one can find an ambiguity in treating the phase of the complex function accompanied by the overall sign ambiguity, which seems closely related to the non-analytic features of the 2D qDRF. To avoid this ambiguity, $F(asb)$ should be written in a single logarithmic term as given in Eq.(\ref{F function single}).

These non-analytic features of the 2D qDRF and the mathematical trick are main results that one may find useful in investigating some characteristic phenomena in 2D and layered electronic systems.

\begin{acknowledgments}
The main contents in this paper comes from the Ch.6 and Ch.7 of C.J. Lee's Ph.D. dissertation(Boston College, 2008). This work was supported by the National Science Foundation under Grant No. PHY-0514619 and PHY-0715227. A refinement in the subsection \ref{discon} and \ref{ambiguity} and also the proofs in the appendix \ref{app long w}, \ref{app col} and \ref{app am} have been done at Donostia International Physics Center(DIPC) in San Sebastian. The contents in this paper are deeply indebted to J. Martin Rommel's Ph.D. thesis.
The author gratefully acknowledges Gabor J. Kalman, J. Martin Rommel and J.M. Pitarke for their careful reading of the manuscript and the useful comments, and also P.M. Echenique for his kind hospitality and the useful comments, and finally special thanks to M. Howard Lee for the warm encouragement.\\
While this paper was under review,  the valuable discussions with N.H. March and I. Nagy motivated me to make brief comments on Ref.\cite{Z91}. I. Nagy had informed me of Ref.\cite{B83} that lead me to add a  new paragraph on the validity of the Eq.(\ref{q comp}).
\end{acknowledgments}

\appendix

\section{\label{app qDRF}Retarded qDRF}

The retarded qDRF $ \chi(\mathbf{k}_1, \omega_1 ; \mathbf{k}_2,\omega_2)$ of a homogeneous electron gas in an arbitrary dimension can be written as follows
\begin{eqnarray}
&& \chi(\mathbf{k}_1, \omega_1 ; \mathbf{k}_2,\omega_2) \nonumber \\
&& =   \frac{1}{2\hbar^2}
   \Big\{ \hat{D}(\mathbf{k}_1, \omega_1 ; \mathbf{k}_2,\omega_2)
  + \hat{D}(\mathbf{k}_2, \omega_2 ; \mathbf{k}_1,\omega_1)\Big\}, \label{chi d}
\end{eqnarray}
where $\hat{D} (\mathbf{k}_1, \omega_1 ; \mathbf{k}_2,\omega_2)$ is the Fourier transform of 
\begin{eqnarray}
&&  \hat{D}(\mathbf{s}_1, \tau_1 ; \mathbf{s}_2,\tau_2) \equiv (-i)^2 
 \theta(\tau_1)\theta(\tau_2)\theta(\tau_2 - \tau_1) \nonumber \\
&& \times  \big{\langle} \Psi_0 \big{|} \big[ \big[
 \hat{\rho}(0,0),\hat{\rho}(-\mathbf{s}_1, -\tau_1) \big] ,
   \; \hat{\rho}(-\mathbf{s}_2, -\tau_2) \big]\big{|} \Psi_0 \big{\rangle}, \phantom{a} \label{D12 st}
\end{eqnarray}
where $\mathbf{s}_a = \mathbf{r} - \mathbf{r}_a$ and $\tau_a = t - t_a$ represent a position vector and a time variable respectively for $a = 1,2$. One can find that $\hat{D} (\mathbf{k}_1, \omega_1 ;
\mathbf{k}_2,\omega_2)$ can be written as
\begin{eqnarray}
  &&\hat{D}(\mathbf{k}_1, \omega_1 ; \mathbf{k}_2,\omega_2) \nonumber \\
  && = \frac{1}{\mathcal{V}} \sum_{\mathbf{k}'} \sum_{l, m}
  \Bigg\{
   \frac{
   \langle 0 |\hat{\rho} (\mathbf{k}')
  |l\rangle \langle l |
  \hat{\rho}^{\dag }(\mathbf{k}_1)
   |m \rangle \langle m |
   \hat{\rho}^{\dag }(\mathbf{k}_2) | 0 \rangle }
   { \big(\omega
   - {\bar{\omega}}_{l0} + i \delta \big)
   \; \big(\omega_2  - {{\bar{\omega}}}_{m0} + i
  \delta \big) } \nonumber \\
  && \hspace{0.5cm} - \frac{
   \langle 0 |
 \hat{\rho}^{\dag}(\mathbf{k}_1)
  |l\rangle \langle l |
  \hat{\rho}(\mathbf{k}')
   |m \rangle \langle m |
   \hat{\rho}^{ \dag}(\mathbf{k}_2)
   | 0 \rangle }
   { \big(\omega
   - {\bar{\omega}}_{m l} + i \delta \big) \; \big(\omega_2 -{{\bar{\omega}}}_{m0}+ i
  \delta \big) } \nonumber \\
   && \hspace{0.5cm} - \frac{ \langle 0 |
   \hat{\rho}^{\dag}(\mathbf{k}_2)
  |l\rangle \langle l |
  \hat{\rho}(\mathbf{k}')
   |m \rangle \langle m |
   \hat{\rho}^{\dag}(\mathbf{k}_1)
   | 0 \rangle }
   {\big(\omega
   - {\bar{\omega}}_{m l} + i \delta \big)
   \; \big(\omega_2  - {\bar{\omega}}_{0l} + i
  \delta \big) } \nonumber \\
  && \hspace{0.5cm} + \frac{\langle 0 |
   \hat{\rho}^{\dag}(\mathbf{k}_2)
  |l\rangle \langle l |
  \hat{\rho}^{\dag}(\mathbf{k}_1)
   |m \rangle \langle m |
   \hat{\rho}(\mathbf{k}')
   | 0 \rangle }
   { \big(\omega
   - {\bar{\omega}}_{0 m}+ i \delta \big) \;  \big(\omega_2 -{\bar{\omega}}_{0l} + i
  \delta \big) }  \Bigg\}, \label{D12}
\end{eqnarray}
where $\delta$ is an infinitesimal positive real value. $|m\rangle$ means a particle-hole pair state that can be written as  $|m \rangle = \big| \tiny{ \begin{array}{l} p \\ h   \end{array}}
\big{\rangle} = \big| \tiny{ \begin{array}{l} \mathbf{q} + \mathbf{k} \\
    \mathbf{q} \end{array} } \big{\rangle}$ with $\epsilon_m =
\epsilon_{\mathbf{q} + \mathbf{k}} - \epsilon_\mathbf{q}$ and
$ \epsilon_\mathbf{q} = \hbar^2 \mathbf{q}^2 / (2m) $. $\hbar
{\bar{\omega}}_{lm}$ is the difference of two single pair excitation
energy $\epsilon_l$ and $\epsilon_m$, i.e., $\hbar {\bar{\omega}}_{lm} = \epsilon_l - \epsilon_m$.
While performing the Fourier transform of Eq.(\ref{rho2}), one can use $ \hat{H}_e(t) = \frac{1}{\mathcal{V}}
\sum_{\mathbf{k}} \Phi(\mathbf{k},t) \hat{\rho}_{\mathbf{k}}^\dag$  and
$\hat{\rho}_{\mathbf{k}} = \sum_{\mathbf{q},\sigma}
\hat{c}_{\mathbf{q},\sigma}^\dag \hat{c}_{\mathbf{q} + \mathbf{k}, \sigma} $ in the interaction picture
with the spin index $\sigma$ and the volume factor $\mathcal{V}$. 

Here it should be emphasized that the infinitesimal real $\delta$ is always positive at this stage as the correct reflection of the retarded effect. However, the sign of the imaginary infinitesimal in Table \ref{Ff} can be either positive or negative because the sign in front of each term in Eq.(\ref{D12}) has been modified so as to make the sign absorbed into its denominators for the purpose of expressing $\chi$ more systematically within the particle-hole symmetry\cite{FN41}.

\section{\label{app srf}Square root complex function}

\begin{figure}
\centering
  \includegraphics[height=0.27\textwidth]{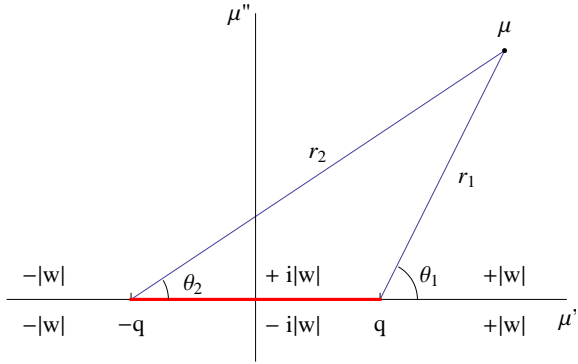}\\
  \caption{(Color online) $w(\mu, q)$ in the vicinity of the real axis in the $\mu$-plane. The thick(red) line is a branch cut.}\label{wz}
\end{figure}

In order to decide which pole is inside the contour, it is crucial to choose the proper form of the complex square root function of $w(\mu, q) \equiv \sqrt{\mu^2 - q^2}$. As a matter of fact, the function of $w(\mu, q)$ can be written in a different form depending on where the branch cut is located as well as which Riemann sheet $\mu$ reside on\cite{Wak}. Let us focus on only the case of $\mu = \mu' + i \mu''$ that is located in the vicinity of the real axis. As for the branch cut, we choose the line connecting $-q$ and $q$ on the real axis as shown in Fig.(\ref{wz}). One can find that $w(\mu, q)$ is purely real in the domain of $|\mu'| > q$ whereas it is purely imaginary in the domain of $|\mu'| \le q$. It also changes the sign as indicated below
\begin{equation}
\sqrt{\mu^2 - q^2}
 =  \left\{
\begin{array}{ll}
 sign(\mu') \sqrt{{|\mu'|}^2 - q^2} \qquad &\textrm{for} \; \;  |\mu'| > q \\
 sign(\mu'') i \sqrt{q^2 - {|\mu'|}^2} \qquad &\textrm{for} \; \; | \mu' | \le q
\end{array} \right. \! \! . \label{sqrt F 3}
\end{equation}
Here it should be emphasized that $w(\mu, q)$ is not well defined at the branch points of $-q$ and $q$. For example, even if $w(\mu, q)$ is continuous at $|\mu'| = q$ as zero, it is not differentiable at the branch point of $q$. However we will treat $w(\mu, q)$ as if it is a well defined function because it is continuous at the branch points of $-q$ and $q$. 

It is noteworthy that the current computational algorithm fails to handle with $w(\mu, q)$ in a correct manner unless $w(\mu, q)$ is defined by Eq.(\ref{sqrt F 3}). 

On the other hand, Rommel and Kalman\cite{RKG98, R99} used 
\begin{equation}
\sqrt{\mu^2 - q^2} = i sign(\mu') \sqrt{q^2 - \mu^2}
\label{RK w}
\end{equation}
where $\mu$ is the complex value of $\mu = \mu' + i \delta$ with the infinitesimal real value of $\delta$ that is always positive.
%

\section{\label{app poles}Poles inside the contour}
A simple way to see which pole is inside the contour is to expand the poles in a Taylor series around the branch point of $q$. 
Let us consider three different cases depending on whether $|\mu'|$ is less than, or greater than, or equal to $q$.
In case of $|\mu'| < q$ we get an approximation of
\begin{equation}
\left|  \frac{z_{\mu}^\pm}{ b} \right| 
 \approx 
1 \pm \frac{1}{2}|\mu''| \sqrt{1 - \big(|\mu'|/q \big)^2} 
\end{equation}
whereas for the case of $|\mu'| > q$ we get
\begin{equation}
\left|  \frac{z_{\mu}^\pm}{ b} \right| 
  \approx    \frac{|\mu'|}{q}
\left\{ 1 + \left[ 1 - (q/|\mu'|)^2 \right] \pm \sqrt{ 1 - \big(q/|\mu'| \big)^2} \right\}.
\end{equation}
For any case above one can find that $z_{\mu}^{+}$ is always outside the contour regardless of the sign and value of
$\mu'$. On the other hand, $z_{\mu}^{-}$ is always inside the contour no matter what
value and sign $\mu'$ has\cite{FN42}. This result does not depend on the sign of $\mu''$ either. At the branch point when $|\mu'| = q$, one may find that both of $z_\mu^\pm/b|$ are outside the contour.  However, this is misleading because we intentionally ignored the small imaginary part of $\mu''$ for the sake of the simplicity when we defined $w(\mu, q)$. As a matter of fact it behaves like $\left| z_{\mu}^\pm / b \right| \approx 1 \pm  |\mu''|/|\mu'|$. 

On the other hand, if one use the Eq.(\ref{RK w}) for $w(\mu,p)$, the pole inside the contour could be either $z_\mu^+$ or $z_\mu^-$ depending on the sign of $\mu'$.

\section{\label{ssB}Alternative way of looking at the vanishing region}

\begin{figure}
\centering
  \includegraphics[height=0.28\textwidth]{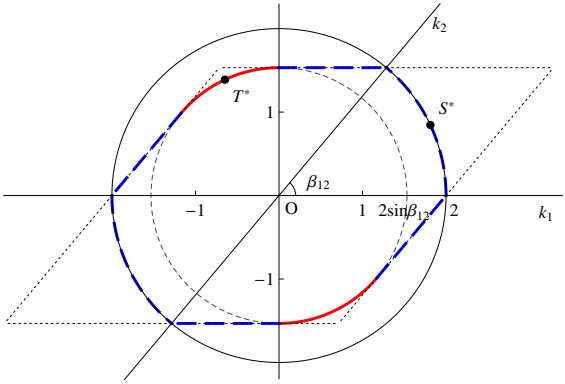}\\
  \caption{(Color online) The vanishing region in contravariant oblique coordinate
    system. The negative axis of $k_1$($k_2$) represents the
    inversion of wavevector $\mathbf{k}_1$($\mathbf{k}_2$).}\label{vanishcontra}
\end{figure}

Interestingly there is an intuitive way to specify the
vanishing region in a direct geometrical manner. Let us consider an
oblique coordinate system in 2D with the angle $\beta$ between $k_1$ and $k_2$
axes as shown in Fig.(\ref{vanishcontra}).

An arbitrary point on the Oblique Coordinate System($OCS$) will be denoted
by $(\zeta, \xi)_O$. This point corresponds to $(\zeta, \xi)_C$ in Cartesian
Coordinate System ($CCS$) through
\begin{equation}
  (\zeta, \xi)_O = (\zeta, \xi)_C,
\end{equation}
which means that they are identical in both the coordinate systems.
However, the distance from the origin is not invariant in the coordinate transformation
from $CCS$ to $OCS$.  

The distance from the origin in $CCS$
is defined as $({\zeta}^2 + {\xi}^2)^{1/2}$ while it is defined by 
$(\zeta^2 + 2 \zeta \xi \cos \beta_{12} + \xi^2)^{1/2}$  in the contra-variant $OCS$.  

Let us make the coordinate transformation of $CCS$ into $OCS$ by rotating the vertical $k_2$ axis in $CCS$ clockwise ending up with $OCS$ in Fig.\ref{vanishcontra}. 
What does the geometric object in $CCS$ transform into in $OCS$? 
One can find the ellipse
$|\eta'|=1$ in Fig.(\ref{Qvanish}) transforms into the circle with radius 2 in $OCS$ as shown in Fig.(\ref{vanishcontra}). 
Besides, the lines of $|\eta_1'| = 1$ and $|\eta_2'|=1$ in $CCS$
transform into the lines that are parallel to the oblique coordinate axes. 
The inner ellipse  $\varphi^+\varphi^- = 0$ transforms into the circle with a
radius $2 \sin \beta_{12}$ inscribed on the parallelogram made out
of $|\eta_1'| = 1$ and $|\eta_2'|=1$.

From the transformation of $CCS$ to $OCS$, one can easily find useful information shared in common by the geometrical objects  in both coordinates.
Firstly, the ratio of radii of two circles ($|\eta'| =1$ and
$\varphi^+ \varphi^- =0$) in $OCS$ is $1:\sin \beta_{12}$, which is exactly the
same as the ratio of semi-axis of two corresponding ellipses in $CCS$.
Secondly, the distance of $OS^*$ in Fig.(\ref{vanishcontra}) is $2$ because
it is on the circle of radius $2$. From this fact, one can specify the
components of $S^* = (\zeta_1, \xi_1)_O$ with $\xi_1 =
\zeta_1$ in $OCS$. Then, by setting $(\zeta_1^2 + 2 \zeta_1 \xi_1 \cos \beta_{12}
+ \xi_1^2)^{1/2} = 2$ one can find $\zeta_1 =
\sqrt{2}/\sqrt{1 + \cos \beta_{12}}$ in both $OCS$ and
$CCS$. The components of $T^* = (-\zeta_2, \xi_2)$ can be obtained
 in the same manner. From the distance of $OT^* = 2 \sin \beta_{12}$ 
and $\beta \equiv \pi - \beta_{12}$ between
$-\mathbf{k}_1$ and $\mathbf{k}_2$, one can find $\zeta_2 = \sqrt{2} \sin \beta
/\sqrt{1 + \cos \beta}$. 

\section{\label{app triangle}Geometrical proof of triangle rule in oblique coordinates}

Here we proove the triangle rule using a simple geometry.
Hereafter the symbol of $\beta_{12}$ will be considered
as an acute angle between $\mathbf{k}_1$ and $\mathbf{k}_2$ while $\beta = \pi - \beta_{12}$
will be used for the obtuse angle between $-\mathbf{k}_1$ and $\mathbf{k}_2$.

Firstly, let us consider a triangle composed of three wave-vector
$\mathbf{k}_1, \mathbf{k}_2$ and $-\mathbf{k}_0$ with $\beta_{12} (< \pi/2)$
and $\mathbf{k}_0 = \mathbf{k}_1 + \mathbf{k}_2$. If it fits inside the boundary in
right upper quadrant enclosed by a bold dashed line in
Fig.(\ref{vanishcontra}), the static 2D $\chi$ vanishes.

Secondly, let us consider the case that the wave-vector $\mathbf{k}_1$ is
inverted to $-\mathbf{k}_1$. Since $\beta = \pi -\beta_{12}$ is greater
than $\pi/2$, one should consider whether or not the triangle fits inside the left upper
quadrant enclosed by the bold and bold-dashed lines in
Fig.(\ref{vanishcontra}). In this case, the inner circle with radius
$2\sin \beta_{12}$ plays an important role since it corresponds to the
condition of $\varphi^+\varphi^- =0$ even though either $k_1$,
$k_2$ or $k$ is not greater than $2k_F$ at all. This condition
actually limits the value of $k$ when $k_1$, $k_2$ and $\beta >\pi/2$ are fixed.

\begin{figure}
\centering
  \includegraphics[height=.23\textwidth]{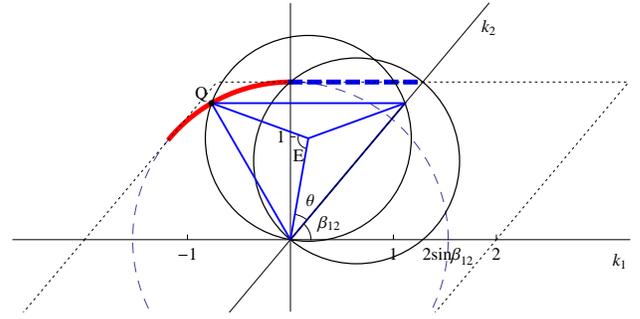}\\
  \caption{\label{triangle}(Color online) Oblique coordinates $(k_1, k_2)$ with the angle $\beta_{12}$. 
The thick(red) arc is the discontinuous vanishing boundary in the oblique coordinates while a solid circle 
stands for a 2D Fermi circle.}
   \vspace{-0.2cm}
\end{figure}

For clarity, let us look at Fig.(\ref{vanishcontra}). Imagine a Fermi circle with radius $1$ 
and put it on Fig.(\ref{vanishcontra}) in a way as its diameter 
coincide with the positive $k_2$-axis in OCS as shown in Fig.(\ref{triangle}). 
Then the center of the Fermi circle will be at $(k_1, k_2)_O =
(0,1)$ along the $k_2$-axis.  The Fermi
circle meets with the line of $k_2 = 2$ at the point where the
inner circle meets the line $k_2 = 2$.

Now let us show how the discontinuous vanishing boundary in Fig.(\ref{vanishcontra}) 
can be constructed by the triangle rule using a simple geometry.
The proof takes three steps. Firstly, we start with the unit Fermi circle whose center is located at
$(0, 1)_O$ as shown in Fig.(\ref{triangle}). Secondly, we rotate the Fermi circle anti-clockwise
by the angle of $\theta$ around the origin. Finally, we draw a straight line parallel to the $k_1$-axis as
indicated by a solid(blue) line in Fig.(\ref{triangle}). The $Q$-point intersecting the
rotated Fermi circle and the solid(blue) line becomes to be on the discontinuous boundary. In addition, it is
not hard to show that the distance of $OQ$ is just $2\sin \beta_{12}$ because the
angle $\angle E$ is $2\beta_{12}$ from a simple geometry. 

In general, the four conditions in Eq.(\ref{k con}) and Eq.(\ref{ex con}) amount to stating
 that the triangle should fit inside the Fermi circle to make 2D static $\chi'$ vanish. 
This feature is called \textit{triangle rule} and was correctly pointed out by Rommel and Kalman\cite{R99}.
%

\section{\label{app long w} Static longwavelength limit}
The vanishing of $\chi (\mathbf{k}_1, \mathbf{k}_2 )$ in the longwavelength limit in Eq.(\ref{q comp}) can be verified in a direct manner. To this end, it suffices to add all six $F$-functions written in a Taylor series as below
\begin{eqnarray}
&& F(\mathbf{k}_a, \mu_a; \mathbf{k}_b, \nu_b)
 \approx  
	\frac{i}{\pi \Delta_{ab}}
	\Bigg\{  - \ln \big( -e^{ - i \beta_{ab}} \big)  \nonumber \\
&&
	+	\log \left(  \frac{a_0}{b_0}  \right) 
 - \left( \frac{a_1}{a_0} + \frac{1}{2 b_0} \right) k_a 
+ \left(  \frac{a_1}{b_0} + \frac{1}{2 a_0} \right) k_b 
	\Bigg\}, \phantom{aaa}
\label{F-function 2}
\end{eqnarray}
where $a_0, b_0$ and $a_1$ are constants defined by $a_0 = - \sin (\beta_{ab}) \textrm{sign}(\mu_a'')$, $b_0 = - \sin (-\beta_{ab}) \textrm{sign}(\nu_b'')$, and $a_1 = \cos (\beta_{ab}) / 2$. The first and second terms contribute to $F'(asb)$ while the other terms contribute to $F''(asb)$. Even if $F'(asb)$ and $F''(asb)$ do not vanish independently, $S'[012]$(or $S'[021]$) vanishes due to the same vanishing mechanism as shown in the $A,B$ column of Table \ref{phase}.

Eq.(\ref{F-function 2}) is general in a sense that it is valid for any angle of $\beta_{12}$ at the static longwavelength limit as $\mathbf{k}_1$ and $\mathbf{k}_2$ goes to zero. 

As a special case of it, let us take a look at the long wavelength limit of Eq.(\ref{collinear}) that is a static collinear-limit expression as $\beta_{12}$ goes to $0$. The vanishing of the right hand side of Eq.(\ref{collinear}) can be verified in a direct manner. To this end, perform a Taylor series expansion of the linear DRF\cite{Lee08b} $\chi(\mathbf{k}_a, \omega_a)$ in Eq.(\ref{linear DRF}), then we get\cite{FN43}
\begin{equation}
\chi(\mathbf{k}_a, \omega_a) 
 \approx 
	- \frac{1}{2 \pi} 
	\left[
	1 + i  \frac{\omega_a'}{2k_a}  \Big( 1 + \frac{k_a^2}{8} + \cdots \Big)
	\right] . \label{linear asym}
\end{equation}
The right hand side of Eq.(\ref{collinear}) can be shown to vanish because of the constancy of $\chi (\mathbf{k}_a, 0) = - 1/( 2 \pi )$ for $k_a \le 2 k_F \; (a = 0,1,2)$ and  the wavevector conservation of $k_0 = k_1 + k_2$.

\section{\label{app col} Collinear limit}

Let us explain how one can reach the static collinear-limit expression of Eq.(\ref{collinear}) from Eq.(\ref{F function single}).
In the parallel collinear limit($\mathbf{k}_1 \parallel \mathbf{k}_2$), we get $\beta_{01} =  \beta_{02} = \beta_{12} \sim \delta \rightarrow 0$ from $\mathbf{k}_0 = \mathbf{k}_1 + \mathbf{k}_2$ and $k_0 = k_1 + k_2$, where $\delta$ is a positive infinitesimal. It is important to find the lowest order term in the asymptotic series of $\cos \beta_{ab}$ and $\sin \beta_{ab}$ for each $F(asb)$ in Table \ref{Ff}. The results are listed in Table \ref{col pap} from which  the asymptotic expressions of $\varphi(\mu_a, \nu_b, + \beta_{ab})$ and $\varphi(\nu_b, \mu_a, - \beta_{ab})$ can be constructed. 
\begin{table}
\caption{\label{col pap} The lowest order term in the asymptotic series of $\cos \beta_{ab}$ and $\sin \beta_{ab}$ as $F(asb)$'s go to the collinear limits.}
\begin{ruledtabular}
\centering
\begin{tabular}{l|cc|cc|cc}
 \multirow{2}{*}{$F(asb)$} &\multicolumn{2}{c}{$\beta_{ab}$} &\multicolumn{2}{c}{$\cos \beta_{ab}$} &\multicolumn{2}{c}{$\sin \beta_{ab}$} \\ \cline{2-7}
 & p\footnote{$p$ column stands for the parallel collinear limit($\mathbf{k}_1 \parallel \mathbf{k}_2$).} & ap\footnote{$ap$ column stands for the anti-parallel collinear limit($\mathbf{k}_1 \parallel -\mathbf{k}_2$) with $k_1 > k_2$. The case of $k_2 > k_1$ is equivalent to melely interchanging the subscripts of $k_1$ and $k_2$.} & p & ap & p & ap \\
\hline
$F(012)$ & $+\delta$                  & $\pi - \delta$              & $+1$  & $-1$          & $+\delta$  & $+\delta$    \\
$F(201)$ & $\pi - \delta$           & $+\delta$                     & $-1$  & $+1$          & $+\delta$   & $+\delta$   \\
$F(120)$ & $+\delta$                  & $+\delta$                     & $+1$  & $+1$          & $+\delta$   & $+\delta$   \\ \cline{2-7}
$F(021)$ & $-\delta$                  & $-\delta$                    & $+1$    & $+1$        & $-\delta$    & $-\delta$  \\
$F(102)$ & $-(\pi - \delta)$       & $-\delta$                    & $-1$   & $+1$        & $-\delta$     & $-\delta$  \\
$F(210)$ & $-\delta$                  & $-(\pi - \delta)$         & $+1$   & $-1$        & $-\delta$     & $-\delta$  \\
\end{tabular}
\end{ruledtabular}
\end{table}

For example, let us focus on $F(012)$. From the $p$-column in Table \ref{col pap}, we know that $\cos \beta_{02} \approx +1$ and $\sin \beta_{02} \approx + \delta$. Then, we can write Eq.(\ref{F function single}) for $F(012)$ as
\begin{equation}
\frac{i}{\pi \Delta_{02}} \ln \left\{ - e^{i \delta} \left( \frac{1- i \delta \xi }{ - ( 1- i \delta \zeta )} \right) \right\}  \approx \frac{i}{\pi \Delta_{02}} \ln \left(  \frac{ 1- i \delta \xi }{ 1- i \delta \zeta }  \right),
\end{equation}
where $\xi \equiv \sqrt{{\eta_0'^+}^2 -1} / ({\eta_0'^+} - {\eta_2'^+} )$ with ${\eta_0'^+} = - k_0/2$ and ${\eta_2'^+} = - k_2 / 2$ whereas $\zeta \equiv  \sqrt{{\eta_2'^+}^2 -1} / ( {\eta_0'^+} - {\eta_2'^+} )$. With $\Delta_{02} \approx k_0 k_2 \delta$, one can find
\begin{eqnarray}
F(012) 
& = & \frac{i}{\pi k_0 k_2 \delta} \big[ \ln ( 1- i \delta \xi ) - \ln (1- i \delta \zeta) \big] \nonumber \\
& \approx & \frac{i}{\pi k_0 k_2 \delta} \big[ - i \delta ( \xi - \zeta ) \big]  \nonumber \\
& = & - \frac{2}{\pi k_0 k_1 k_2} \left\{ \sqrt{{\eta_0'^+}^2 -1} - \sqrt{{\eta_2'^+}^2 -1}   \right\}, \phantom{a}
\end{eqnarray}
where $\ln(1 + x) \approx x$ and $k_0 = k_1 + k_2$ have been used. In a similar way, one can get the other $F(asb)$'s. Using the 2D linear DRF of Eq.(35) in Ref.\cite{Lee08b} below
\begin{equation}
\chi (\mathbf{k}_a, \omega_a) = - \frac{1}{2 \pi k_a} \left\{ k_a +  \sqrt{{\eta_a^+}^2 - 1} + \sqrt{{\eta_a^-}^2 - 1} \right\}, \label{linear DRF}
\end{equation}
one can use the static $\chi (\mathbf{k}_a, 0)$ to reach the static collinear-limit expression of Eq.(\ref{collinear}). As for the anti-parallel collinear limit, refer to $ap$-column in Table \ref{col pap} where the wavevector conservation of $k_0 = k_1 - k_2$ is implied on the assumption of $k_1 > k_2$.

\section{\label{app am} Phase ambiguity}

Here it is explained how the phase ambiguity comes about. Let us focus on the $4 \times 4$ square in Cartesian coordinates $(k_1, k_2)_C$ and consider the values of $\vartheta_1[012]$ and $\vartheta_2[012]$. In the second case, we find $\vartheta_1[012] = 2 \pi$ and $\vartheta_2[012] = -2\pi$, so that $\vartheta[012] = 2\pi - 2\pi = 0$ in the entire region from A to E, which means the vanishing extends even into the region C without the discontinuous boundary of $\varphi^+ \varphi^- = 0$. Here one should take it more serious to violate the triangle rule than to find no discontinuity.

There is an alternative way to look at the extension of the vanishing region in the second case. From the fact of $\ln(e^{i \vartheta_1[012]}) = \ln(e^{i 2\pi}) = \ln 1 = 0$, each phase-sum of $\vartheta_1[012]$ and $\vartheta_2[012]$ is equivalent to $0$ or congruent to 0. In other words they vanish independently. Therefore, in the second way to sum up the phases, the term of $\vartheta_2[012]$ always vanishes with no effect at all, so that $\chi'$ can be determined solely by $\vartheta_1[012]$.

Unfortunately this ambiguity brings on the overall sign ambiguity accompanied by the value of $\vartheta[012]$ itself changed. 
To make it clearer, let us focus on the term $\vartheta_1(asb)$ in $S[012]$ ignoring $\vartheta_2[012]$ because it vanishes in the second case. We find that any $\vartheta_1(asb)$ is not less than 0, so that $\vartheta[012]$ is also positive making $\chi'$ be negative. However, this contradicts the plot of $\chi'$ obtained from the first case, where the sign of $\chi'$ is opposite. 

Where does the sign ambiguity come from? 
To answer the question, we should go back to the first case. Let us look at the effect of $\vartheta_2(asb)$ on $\vartheta(asb)$. We find any $\vartheta_2(asb) = -(\pi - \beta_{ab})$ is not greater than 0 because of $0< \beta_{ab} < \pi$ in $S[012]$.  On the other hand, the phase of any $\vartheta_1(asb)$ is in the range of $0 \le \vartheta_1(asb) \le \pi$. Then, we find $ -(\pi -\beta_{ab}) \le \vartheta(asb) \le \pi - (\pi - \beta_{ab})$. Now let us recall the next step taken in the first case. We should sum up the three $\vartheta(asb)$'s to get $\vartheta[012]$, which makes three $\beta_{ab}$'s cancel each other out due to the fact of $\beta_{12} = \beta_{10} + \beta_{02}$. If this cancellation was taken into account in advance, the phase of $\vartheta(asb)$ would be in the range of $ -\pi  \le \vartheta(asb) \le 0$. Compare it with the phase of $0 \le \vartheta_1(asb) \le \pi$ that are only terms counted on in the second case. We find the sign of the former is opposite the latter. Which is the correct sign?

We do not find any reason to violate the \textit{triangle rule} by separating two logarithmic terms of $F(asb)$, so that the plots of $\chi'$ in Fig.(\ref{fig:3d}) look feasible even if there is an overall sign ambiguity. Moreover, the second way to sum up the phase is equivalent to removing the second logarithmic term of $F(asb)$ in Eq.(\ref{F function 1}) or Eq.(\ref{F function}). However, this removal can lead $F(asb)$ to diverge for $\beta_{12} = 0, \pi$ in the static short-wavelength limit or in the high-frequency limit. To avoid the ambiguity and the divergence in a numerical calculation, we should write $F(asb)$ in a single logarithmic term as given in Eq.(\ref{F function single}).


\begin{references}
\bibitem{FN24} For example, if one plots $F(201)$ with $k_2 = 2 k_1$ and $\beta_{12} = 3\pi/4$ in Mathematica, one may find a certain region where $F(201)$ discontinuously changes its sign twice. It corresponds to the region D(or similarly E) in Fig.(\ref{Qvanish}).
\bibitem{FN25} Correction: The corresponding expression in page 60 of Ref.\cite{Lee08a} should be replaced by Eq.(\ref{rho2}) in this paper. Eq.(\ref{rho2}) is identical with Eq.(2.3) in BPE except the overall sign.
\bibitem{FN28} It is also assumed that $z_{\mu_a}^-$ and $z_{\nu_b}^-$ can not coincide with each other to remain as a simple pole.
\bibitem{FN29}The complex function of Eq.(\ref{srf}) is not well-defined at $\mu_a' = q$. However, it will be treated as if it is defined at $\mu_a' = q$ because it is continuous at $\mu_a' = q$.
\bibitem{FN30} Eq.(\ref{sym F3}) can be verified from $\eta_a^- = (\eta_a^+)^*$ with $\eta_a^\pm = \pm i \delta - k_a / 2$.
\bibitem{FN31} As for the static case, the exclusion condition in Eq.(\ref{ex con}) can be specified by $\varphi^+ \varphi^- = {\eta_1'}^2 - 2 \eta_1' \eta_2' \cos \beta_{12} + {\eta_2'}^2 - \sin^2 \beta_{12}$ and $\ell^+ \ell^- =  (\eta_1' \cos \beta_{12} - \eta_2')( \eta_2' \cos \beta_{12} - \eta_1' )$.
\bibitem{FN32} The distance of $OT^*$ should not be confused with the length of $\mathbf{k}_0$. Actually as $\beta_{12}$ goes to $\pi$, $k_0$ goes to zero in Oblique coordinates $(\mathbf{k}_1, \mathbf{k}_2)_O$ whereas $OT^*$ goes to $2\sqrt{2}$ in Cartesian coordinates $(k_1, k_2)_C$.
\bibitem{FN35} The proof is simple. Assume $k_1 \ge k_2$. Then we know $\chi'(\mathbf{k}_0) \ge \chi'(\mathbf{k}_1) \ge \chi' (\mathbf{k}_2)$ from the well-known 2D $\ell$DRF. Then we find that $\left[ k_0  \chi' (\mathbf{k}_0 \! ) - k_1 \chi' (\mathbf{k}_1 \! ) - k_2 \chi' (\mathbf{k}_2 \!) \right] \ge \left[ k_0  \chi' (\mathbf{k}_0 \! ) - k_1 \chi' (\mathbf{k}_1 \! ) - k_2 \chi' (\mathbf{k}_1 \!) \right] \ge k_0 \left[ \chi'(\mathbf{k}_0) - \chi'(\mathbf{k}_1) \right] \ge 0$.  Therefore, $\chi' (\mathbf{k}_1; \mathbf{k}_2 ) \ge 0$ in the parallel collinear limit.
\bibitem{FN36} Correction: The value of $\vartheta[012]$ in the column C of the Table7.2 in Ref.\cite{Lee08a} should be corrected from $\pi$ to $0$. Moreover, in page 93-94 of Ref.\cite{Lee08a}, the vanishing mechanism based on  the phase congruence is only partially successful, however, bringing on an ambiguity in the way to sum the phase of $\vartheta(asb)$.\\ Caution: The phase $\vartheta$ in this paper is different from the same notation $\vartheta$ in Ref.\cite{Lee08a}, where $\vartheta$ denotes only the phase of $\ln(\varphi^+/\varphi^-)$.
\bibitem{FN39} The single logarithmic term expression of Eq.(\ref{F function single}) is equivalent to  Eq.(\ref{F function 1}) too, and it shows the correct property of the evenness with respect to $\beta_{ab}$ for the visualization of 2D qDRF.
\bibitem{FN40} The form of Eq.(\ref{F function single}) fits the conventional phase congruent region of $(-\pi, \pi]$ adopted by a computational algorithm. For example, as for $F(012)$, the phase of the logarithmic term in Eq.(\ref{F function single}) should be $\vartheta(012) = (\pi + \beta_{02}) + \vartheta_1(012)$ from the analytic point of view. However, this phase may be treated as $-(\pi - \beta_{02}) + \vartheta_1(012)$ by a computational algorithm through the following two steps: convert $(\pi + \beta_{02})$ into $-(\pi - \beta_{02})$, and then add it to $\vartheta_1(012)$. Actually the treatment of the phase in a computational algorithm seems to coincide with our analysis in the subsection \ref{discon}. However it is just a naive guess.  A Mathematica code for the visualization of 2D qDRF can be found in Ref.\cite{Lee08a}.  For details, refer to the explanation therein.
\bibitem{FN42} As for the case of $|\mu'| > p$, the determination can be made easy from the fact that  $\sqrt{x}$ is always greater than $x$ for $0<x<1$ with $x \equiv 1 - (p/|\mu'|)^2$.
\bibitem{FN43} The factor of $-1/(2 \pi)$ in Eq.(\ref{linear asym}) differs from the well-known value of $-1/\pi$ because of the different energy unit adopted in this paper.
\end{references}
\end{document}